\documentstyle[12pt,fps]{article}
\makeatletter
\@addtoreset{equation}{section}
\makeatother
\renewcommand{\theequation}{\thesection.\arabic{equation}}
\newcommand{\bibi}{\bibitem}

\newcommand{\un}{1\!\!1}

\newcommand{\al}{\alpha}

\newcommand{\bt}{\beta}
\newcommand{\lag}{\langle}
\newcommand{\rag}{\rangle}
\newcommand{\gm}{\gamma}

\newcommand{\dl}{\delta}
\newcommand{\ep}{\varepsilon}

\newcommand{\et}{\eta}
\newcommand{\m}{\mu}
\newcommand{\h}{\eta}

\newcommand{\kp}{\kappa}
\newcommand{\lm}{\lambda}

\newcommand{\ps}{\psi}
\newcommand{\om}{\omega}

\newcommand{\Ps}{\Psi}

\newcommand{\Om}{\Omega}
\newcommand{\Psb}{\overline{\Ps}}
\newcommand{\psb}{\overline{\ps}}

\newcommand{\cp}{\chi^{\prime}}
\newcommand{\cpb}{\overline{\chi}^{\prime}}

\newcommand{\mh}{\hat{\mu}}

\newcommand{\chb}{\overline{\chi}}

\newcommand{\hmu}{\hat{\mu}}

\newcommand{\aleq}{\,\mbox{}^{\textstyle <}_{\textstyle\sim}\,}

\newcommand{\eighth}{\mbox{{\small $\frac{1}{8}$}} }

\newcommand{\Tr}{\mbox{Tr\,}}
\newcommand{\IM}{\mbox{Im\,}}
\newcommand{\RE}{\mbox{Re\,}}
\newcommand{\Det}{\mbox{Det\,}}
\newcommand{\del}{\partial}

\newcommand{\dg}{\dagger}
\newcommand{\pr}{\prime}
\newcommand{\ra}{\rightarrow}

\newcommand{\be}{\begin{equation}}
\newcommand{\ee}{\end{equation}}
\newcommand{\bea}{\begin{eqnarray}}
\newcommand{\eea}{\end{eqnarray}}
\newcommand{\append}{  \setcounter{section}{1}\setcounter{equation}{0}
          \section*{Appendix}
   \renewcommand{\theequation}{\Alph{section}.\arabic{equation}}  }
\newcommand{\eq}{\ref}
\newcommand{\beq}{\begin{equation}}
\newcommand{\eeq}{\end{equation}}
\newcommand{\cc}{\cite}
\newcommand{\lb}{\label}
\newcommand{\gsim}{\stackrel{>}{\sim}}  % >~
\newcommand{\lsim}{\stackrel{<}{\sim}} % <~

\def \3{\ss}

\def\dateandnumber(#1)#2#3#4{
\vbox to 15mm{%
     \hbox to \textwidth{ \hspace*{14mm} \hsize=40mm%
            \vbox{%
                 \hbox to 40mm{\large #1 \hss}%
                 \hbox to 40mm{    \hss}%
                 \hbox to 40mm{    \hss}%
                 }%
                 \hss \hsize=80mm%
            \vbox{%
                 \hbox to 80mm{\hss \large #2}
                 \hbox to 80mm{\hss \large #3}
                 \hbox to 80mm{\hss \large #4}
                 }%
            \hspace*{14mm} }%
      \vss
    }
}
\def\titleofpreprint#1#2#3#4{{\LARGE \bf
\vbox to 40mm{%
     \vss
     \hbox to \textwidth{ \hspace*{14mm} \hsize=130mm%
            \hss \vbox{% \LARGE \bf
                      \hbox to 130mm{\hss \LARGE \bf #1\hss}%
                      \hbox to 130mm{\hss \LARGE \bf #2\hss}%
                      \hbox to 130mm{\hss \LARGE \bf #3\hss}%
                      \hbox to 130mm{\hss \LARGE \bf #4\hss}%
                 }%
            \hss \hspace*{14mm} }%
      \vss
    }}
}
\def\listofauthors#1#2#3{{\large
\vbox to 18mm{%
     \vss
     \hbox to \textwidth{ \hspace*{14mm} \hsize=130mm%
            \hss \vbox{% \large
                      \hbox to 130mm{\hss \large #1\hss}%
                      \hbox to 130mm{\hss \large #2\hss}%
                      \hbox to 130mm{\hss \large #3\hss}%
                 }%
            \hss \hspace*{14mm} }%
      \vss
    }}
}
\def\listofaddresses#1#2#3#4#5{{\small
\vbox to 16mm{%
     \vss
     \hbox to \textwidth{ \hspace*{14mm} \hsize=130mm%
            \hss \vbox{% \small
                      \hbox to 130mm{\hss \small #1\hss}%
                      \hbox to 130mm{\hss \small #2\hss}%
                      \hbox to 130mm{\hss \small #3\hss}%
                      \hbox to 130mm{\hss \small #4\hss}%
                      \hbox to 130mm{\hss \small #5\hss}%
                 }%
            \hss \hspace*{14mm} }%
      \vss
    }}
}
\def\abstractofpreprint#1{{\normalsize
  \vbox to 100mm{%
     \vss
     \hbox to \textwidth{\hss \normalsize \bf Abstract \hss}%
     \vspace*{1cm} \normalsize
     #1
     \vss
    }}
}

\def\footnoteitem(#1)#2{
\begin{list}{#1}{\labelwidth4.0mm \leftmargin7.0mm
\labelsep2.5mm \rightmargin7.0mm \parsep0.5ex plus0.2ex minus0.1ex
\itemsep0ex plus0.2ex }
\item #2
\end{list}
}
\begin{document}
%%%%%%%%%%%%%%%%%%%%%%%%%%%%%%%%%%%%%%%%%%%%%%%%%%%%%%%%%%%%%%%%%
\dateandnumber(August 1993)%
{Amsterdam ITFA 93-18}%
{UCSD/PTH 93-15}%
{              }%
%%%%%%%%%%%%%%%%%%%%%%%%%%%%%%%%%%%%%%%%%%%%%%%%%%%%%%%%%%%%%%%%%
\titleofpreprint%
{      Non-gauge fixing approach to                            }%
{      chiral gauge theories using                             }%
{      staggered fermions                                      }%
{                                                              }%
%%%%%%%%%%%%%%%%%%%%%%%%%%%%%%%%%%%%%%%%%%%%%%%%%%%%%%%%%%%%%%%%%
\listofauthors%
{Wolfgang Bock$^{1,\#}$, %
Jan Smit$^{1,\&}$ and Jeroen C. Vink$^{2,*}$                 }%
{                                                              }%
{                                                              }%
%%%%%%%%%%%%%%%%%%%%%%%%%%%%%%%%%%%%%%%%%%%%%%%%%%%%%%%%%%%%%%%%%
\listofaddresses%
{\em $^1$Institute of Theoretical Physics, University of Amsterdam,}%
{\em \\ Valckenierstraat 65, 1018 XE Amsterdam, The Netherlands }%
{\em $^2$University of California, San Diego, Department of Physics,}%
{\em \\ 9500 Gilman Drive 0319, La Jolla, CA 92093-0319, USA}%
{\em            }%
%%%%%%%%%%%%%%%%%%%%%%%%%%%%%%%%%%%%%%%%%%%%%%%%%%%%%%%%%%%%%%%%%%
\abstractofpreprint{
We investigate a proposal for the construction
of models with chiral fermions on the lattice using
staggered fermions. In this approach the gauge invariance is
broken by the coupling of the staggered fermions to the gauge fields.
Motivated by previous results in the non-gauge invariant massive
Yang-Mills theory and certain gauge-fermion models we
aim at a dynamical restoration of the gauge invariance in the full
quantum model. If the gauge symmetry breaking is not too severe,
this procedure could lead in the continuum limit to the desired gauge
invariant chiral gauge theory.
This scenario is very attractive since it does not rely on gauge fixing.
We investigate a simple realization of
this approach in a U(1) axial-vector model with
dynamical fermions in four dimensions.
}
%%%%%%%%%%%%%%%%%%%%%%%%%%%%%%%%%%%%%%%%%%%%%%%%%%%%%%%%%%%%%%%%%
\nopagebreak
\vspace{-6mm}
 \noindent $\#$ e-mail: bock@phys.uva.nl \\
 \noindent $\&$ e-mail: jsmit@phys.uva.nl \\
 \noindent $*$  e-mail: vink@yukawa.ucsd.edu \\
\pagebreak
\section{Introduction}
An important unsolved problem in lattice field theory is the non-perturbative
formulation of a chiral gauge theory on the lattice
(for
reviews see refs.~\cc{Sm88,REV}).
The naive lattice transcription of a quantum field theory
involving fermions leads to unwanted fermion species, the so-called fermion
doublers.
Half of these couple with opposite chiral charge to the
gauge fields, spoiling the chiral nature of the couplings,
and the resulting theory in the scaling region is vector-like.
To deal with these doublers, one has three options:
I) Remove the fermion doublers
by rendering them heavy, while keeping only one fermion in the
physical spectrum [3-11],
% \cc{WY,EP,ROME,BK,ZK,KA,PR,FS,Ne},
II) decouple them by turning off their interactions with the other
particles \cc{ZA} or
III) use them as physical degrees of freedom \cc{Sm88,Sm91,Sm92}.

When applied to chiral models, one has to reconcile the chosen
regularization with gauge invariance.
Many proposals are formulated such that
local gauge invariance on the lattice is preserved [3,4,7-9,11].
% \cc{WY,EP,ZK,KA,PR,Ne}.
For example,
using a Wilson term
to decouple the doublers, one introduces  extra scalar fields to
make this mass term gauge invariant.
Alternatively one can sacrifice gauge invariance of the lattice model
and transcribe the gauge fixed continuum action to the lattice \cc{ROME,BK}.
This avoids introducing extra scalar fields, but requires
Fadeev-Popov and gauge fixing terms.
It can be done with Wilson fermions as in refs.~\cc{ROME,BK}, but
also with staggered fermions. In either case one breaks the BRST symmetry,
and one has to add counterterms to the action and
tune their coefficients such that BRST invariance is restored in the
continuum limit. This method is cumbersome from a technical point of view
and in addition one has to worry about non-perturbative
gauge fixing.

The method  which we will focus on in this paper falls in class III.
It relies on staggered fermions, which method uses the species doublers as
Dirac-flavor components.
The one-component staggered fermion fields do not carry explicit Dirac
and flavor labels,
these components are `spread out' over the lattice.
In the classical continuum limit this method leads to four
flavors of Dirac fermions.
It is possible to couple these Dirac  flavors to the gauge
fields such that the chiral target model is recovered in the
classical continuum limit.
We studied the method
in ref.~\cc{SMOOTH}
for the case of a two-dimensional model with axial-vector couplings
and concluded that it performed well for smooth external gauge fields.
The important issue, however, is whether
the same can be achieved also when the full quantum
fluctuations are taken into account.
The same method of coupling the staggered
Dirac-flavor components has recently been successfully used
for an investigation of
a strongly coupled fermion-Higgs model \cc{UP}.

In chiral gauge models we are confronted with the problem that the
couplings of the staggered fermions to the gauge fields
spoil the local gauge invariance.
One possibility to overcome this difficulty
is
described in ref.~\cc{ROME}, using the full machinery of
gauge fixing.
Alternatively one can attempt to avoid this and hope for a dynamical
restoration of gauge invariance in the model without gauge fixing.
By
this we mean the following:
carrying out the
integration over all gauge field configurations in the
path integral, the gauge degrees of freedom appear in the action
as dynamical scalar fields with frozen radial mode; when these
scalar fields decouple, gauge invariance is restored.
As we shall see later there exist examples of non-chiral models
where these scalar fields indeed decouple for suitable
choices of the bare parameters.
It is the subject of
this paper to investigate this scenario for the staggered fermion
approach to chiral gauge theories.

The outline of the paper is as follows: In sect. 2 we
recall
the
scenario of dynamical gauge symmetry restoration and
review
the massive Yang-Mills model and a gauge-fermion model, as examples
of models in which this scenario was found to work.
In sect.~3 we define two staggered fermion models
with axial-vector couplings to a U(1) gauge field: the ISF
(invariant staggered fermion) model, which has a local gauge invariance
and the NISF (non-invariant staggered fermion) model,
which  couples the staggered Dirac-flavor components in a way
that allows for the construction of chiral models, but
breaks local gauge invariance.
The phase diagram of the ISF model will be presented in sect.~4.
We shall also break the gauge symmetry of the ISF model by hand and
argue that in this case the gauge invariance
can again be restored
dynamically.
The phase diagram of the NISF model is presented in sect.~5 and from it
we infer that the desired symmetry restoration
does most probably not take place.
We try to improve on this by modifying the model
in sect.~6.
Also here we find no evidence for the dynamical
restoration of the gauge symmetry. Sect.~7 contains a
summary of our results and gives an outlook to possible future
investigations.
\section{Restoration of gauge symmetry}
There exist examples of gauge non-invariant
lattice models where a dynamical restoration of gauge symmetry takes
place.  More precisely it has been found that one can add terms to
a gauge invariant action that break the gauge symmetry, but provided
the bare coupling constants of these terms are not too large, the
low energy model remains the same. In our case the gauge symmetry is
broken by the  lattice regularization with staggered fermions, and
there is no easy separation between a gauge invariant action and additional
gauge symmetry breaking terms. The question arises if this symmetry
breaking is sufficiently small that the chiral
symmetry  gets restored dynamically.

To explain the basic idea
let us start first from a generic non-gauge invariant lattice action
$S_{eff}(U^{\prime})$. If also fermions are involved
$S_{eff}(U^{\prime})$  represents the effective action after
integrating out the fermionic degrees of freedom in the path integral.
The link field $U_{\mu x}=\exp(-iaA_{\mu x})\in G$, with
$a$ the lattice distance, $A_{\mu x}$ the vector potential
and $G$ the compact gauge group.
In the following we shall mostly use lattice units, $a=1$.
Assuming that the partition function is defined by the usual
integration over the link field, it can be written as
\bea
 Z \!\!\!& = &\!\!\! \int D U^{\prime} \exp S_{eff}(U^{\prime}_{\mu x})
 = \int D U \exp S_{eff}(\Om_x U_{\mu x} \Om_{x+\hmu}^{\dg}) \nonumber\\
  \!\!\!& = &\!\!\! \int D U D V \exp S_{eff}(V_x^{\dg} U_{\mu x}
V_{x+\hmu})\;, \lb{PART}
\eea
where we have used in the second equation the gauge transformation
$U^{\prime}_{\mu x}=\Om_x U_{\mu x} \Om^{\dagger}_{x+\hmu}$
as a transformation of variables
as well as the invariance of the Haar measure, $D U^{\prime} =D U$.
In the third equation we have added the trivial integration
$1=\int D \Om$, and wrote $V_x=\Om^{\dagger}_x$. In this way the
gauge degrees of freedom $\Om$ have been exhibited as an
additional field in the path integral, which could be interpreted
as a dynamical Higgs fields $V\in G$ with a frozen radial mode,
$V^{\dagger}_x V_x =\un$.
The new action $S_{eff}(V_x^{\dg} U_{\mu x} V_{x+\hmu})$
is invariant under the local gauge transformations
$U_{\mu x} \ra \Om_x U_{\mu x} \Om^{\dagger}_{x+\hmu}$,
$V_x \ra \Om_x V_x$.

The above manipulations turn any gauge non-invariant model into a
related gauge invariant model, at the price of introducing an extra
scalar field $V_x\in G$. The important question is if the resulting model
still describes the physics of the underlying gauge invariant model, i.e.
the models without adding the symmetry breaking terms to
the action or, in our case, the model of the chirally invariant target
model in the continuum.
For instance, this requires that the $V$ field
decouples
in the scaling region.

Before we turn to the question of dynamical gauge symmetry
restoration in the staggered fermion model, let us first present two
specific examples, namely the massive Yang-Mills theory and certain
gauge-fermion models,
which show that gauge non-invariant terms
with coefficients which are not too large,
do not spoil gauge invariance after the integration over
quantum fluctuations in the path integral has been carried out.
\subsection{Massive Yang-Mills model}
The massive $SU(2)$
Yang-Mills model
is defined on the lattice by the euclidean action,
\be
S_U  =  \frac{2}{g^2} \sum_{x \mu \nu}
\mbox{Tr\,} U_{x \mu \nu} + \kp \sum_{x \mu}
\mbox{Tr} \left\{ U_{\mu x} + U_{\mu x}^{\dg} \right\}
\;, \lb{YM}
\ee
where $g$ is the gauge coupling  and $\kp$
the bare mass parameter in lattice units ($a=1$). The
$U_{x \mu \nu}$ denotes the usual plaquette variable on the lattice.
The mass term in (\eq{YM}) breaks the local SU(2) gauge invariance.
After going through the steps of eq.~(\eq{PART}) we find the action of
the gauge invariant extension of the model,
\be
S_{U,V}= \frac{2}{g^2} \sum_{x \mu \nu}
\mbox{Tr\,} U_{x \mu \nu} + \kp \sum_{x \mu} \mbox{Tr}
\left\{ V^{\dg}_x U_{\mu x} V_{x+\hmu}+ V_{x+\hmu}^{\dg} U_{\mu x}^{\dg}
V_{x}
 \right\} \;,\lb{YMH}
\ee
which is the action of a gauged SU(2)$\times$SU(2) non-linear sigma model.
Note that the perturbative non-renormalizability of the model (\ref{YM})
causes no problems when the model is treated non-perturbatively.

The non-linear sigma model at $g=0$ (i.e. for $U_{\mu x}=\un$)
has a phase transition at $\kp=\kp_c(0) \approx 0.3$.
For $g>0$ this
phase transition extends into the $\kp$-$g$ plane and separates a
Higgs ($\kp>\kp_c(g)$) from a confinement phase ($\kp<\kp_c(g)$).
We are interested here only in the scaling region at small bare
gauge coupling $g$, where we find continuum behavior.

There are three scaling regions of interest:
(A) When approaching the point $\kp_c(0)$ from within
the Higgs phase, we encounter the usual Higgs phenomenon. The
spectrum contains the three massive gauge bosons and the Higgs particle.
(B) When approaching the point $\kp_c(0)$ from within the confinement phase
we  have a theory with confined scalar particles, much like QCD.
(C) If we let $g$ approach zero inside the confinement phase
away  from the Higgs-confinement phase transition $\kp_c(g)$,
the scalar particles acquire masses of the order of
the cut-off and we recover the pure SU(2) Yang-Mills system
with only glueballs in the physical particle spectrum.

In case C there is a whole region in parameter space
($0 < \kp < \kp_c(g)$), where the
physics in the scaling region is the same as at
$\kp=0$, at which point the bare action is that of the gauge invariant
Yang-Mills model. However, if we make the symmetry breaking too large
(i.e. $\kp$ close to $\kp_c$),
the model changes and goes into a different phase in which the physics
is no longer described by the pure Yang-Mills part of (\eq{YM}).
It is also interesting to note in passing that in case (A) the
gauge degrees of freedom $V$ produce a genuine Higgs field, which
acquires a non-frozen radial mode in the low energy action.
The representation of the gauge group carried by this Higgs
field is determined by the initial symmetry breaking.
\subsection{Gauge-fermion models}
As a second example we discuss restoration of gauge invariance in
models with fermions in which the symmetry breaking
resides in the fermionic part of the action.
The
model which we shall consider here, and in more detail in the
next sections, has
axial-vector couplings to U(1) gauge fields,
\be
S = -\int d^4 x
\sum_{f=1}^{N_F} \left\{ \psb_f^\pr \gamma_{\mu}
(\partial_{\mu} + iq_f \gamma_5 A_{\mu}) \psi_f^\pr +
y\psb_f^\pr\psi_f^\pr \right\}\;.
\lb{AXCO}
\ee
There are $N_F$ fermion flavors with charges $q_f=\pm 1$,
$f=1,\ldots,N_F$, such that the model is anomaly free.
For $y=0$ the action (\eq{AXCO}) is invariant
under the local gauge transformation:
$A_{\mu}(x)\ra A_{\mu}(x)+\del_{\mu}\om(x)$,
$\psi(x) \ra (\Om(x) P_L+\Om^*(x)P_R) \psi(x)$,
$\psb(x) \ra \psb(x) (\Om(x) P_L+\Om^*(x)P_R)$,
with
$\Om(x) = \exp i\om(x)$ and
$P_L=(1 - \gm_5)/2$, $P_R=(1 + \gm_5)/2$.
A non-zero mass parameter $y$ allows us to discuss the effect of
gauge symmetry breaking by a fermion mass.
We have put a prime on $\psi$ and $\psb$ to indicate
the gauge in which the symmetry breaking has the simple mass term
form $\sum_f y\psb_f^\pr\psi_f^\pr$.
Our target model is the gauge invariant model at $y=0$. We
shall show that this can be achieved even for nonzero $y$,
provided it is not too large,

The lattice transcription of the action (\eq{AXCO})
with a single naive fermion field reads
\bea
S_{U,\psi^\pr} \!\!\!&=&\!\!\! -\frac{1}{2} \sum_{x \mu} \left\{ \psb_x^\pr
\gm_{\mu}
( U_{\mu x} P_L + U_{\mu x}^*  P_R )
\psi_{x+\hmu}^\pr - \psb_{x+\hmu}^\pr \gm_{\mu}
(U_{\mu x}^* P_L+U_{\mu x} P_R)
\psi_x^\pr \right\} \nonumber \\
\!\!\!& &\!\!\! - y \sum_x \psb_x^\pr \psi_x^\pr \;. \lb{NAIV}
\eea
Because of the species doubling phenomenon this action
reduces in the classical continuum limit to (\eq{AXCO})
with $N_F=16$, where eight of the fermion species couple to the
gauge fields with $q_f=+1$ and the remaining eight with $q_f=-1$.
Note that with these charges the model is anomaly free and
equivalent to QED with 16 flavors.
The model we shall actually study numerically is a staggered
fermion reduction of (\ref{NAIV}) in which the number of flavors
is reduced by a factor of two (four $q_f$'s equal to $+1$ and four
equal to $-1$). This model we shall call in the following the ISF model.
However, in order not to overload
the reader here with the details of the staggered fermion formalism, we shall
continue for the moment with (\ref{NAIV}) as its properties are
qualitatively the same as the ISF model.

For $y=0$ the action (\eq{NAIV})
is invariant under the local gauge transformations,
\be
U_{\mu x} \ra \Om_x U_{\mu x} \Om_{x+\hmu}^* \;,\;\;\;
\psi_x^\pr  \ra (\Om_x P_L+\Om^*_x P_R) \psi_x^\pr   \;,\;\;\;
\psb_x^\pr \ra \psb_x^\pr (\Om_x  P_L+\Om^*_x P_R)   \;. \lb{ET}
\ee
For non-zero $y$
the mass term in the action
breaks gauge invariance. We shall argue however, that
as in the case of the massive Yang-Mills model,
the physics in the
scaling region remains the same as with $y=0$, provided
$y$ is not too large. Our arguments below are based on the
knowledge of Wilson-Yukawa models accumulated in recent years
\cc{PMS}.

Starting from the gauge non-invariant action (\eq{NAIV}) and
after going through similar steps as in (\eq{PART}) we find,
\bea
S_{U,V,\psi^\pr}\!\!\!&=&\!\!\!-\frac{1}{2} \sum_{x \mu}\left\{
\psb^\pr_x \gm_{\mu} ( V_x^* U_{\mu x} V_{x+\hmu} P_L+
  V_{x+\hmu}^* U_{\mu x}^* V_x P_R) \psi^\pr_{x+\hmu}
\right.\nonumber\\
 \!\!\!& &\!\!\!\left. -
\psb^\pr_{x+\hmu} \gm_{\mu} (V_{x+\hmu}^* U_{\mu x}^* V_x P_L
   + V_x^* U_{\mu x} V_{x+\hmu}P_R) \psi^\pr_x \right\}
  - y \sum_x \psb^\pr_x \psi^\pr_x \;.
\lb{YUKPR}
\eea
The fields $\psi^\pr$ and $\psb^\pr$ are now seen to be screened from the
gauge fields by the $V$ fields and therefore are neutral with respect to
the U(1)
gauge transformations
\be
U_{\mu x} \ra \Om_x U_{\mu x} \Om_{x+\hmu}^* \;,\;\;\;
V_x \ra \Om_x V_x \;,\;\;\;
\psi^\pr_x  \ra \psi^\pr_x \;,\;\;\;
\psb^\pr_x \ra \psb^\pr_x. \lb{NEUTRALGT}
\ee
To see
how the symmetry gets restored it is useful
to make a transformation of variables to
new fermionic fields
\be
\psi^\pr_x  = (V_x^* P_L+V_x P_R) \psi_x   \;,\;\;\;
\psb^\pr_x  = \psb_x (V_x^*  P_L+V_x P_R)   \;, \lb{PSPR}
\ee
which leaves the integration measure invariant, $D \psb D \psi=
D \psb^{\prime} D \psi^{\prime}$. The new fields $\psi$ and $\psb$
carry
a U(1) charge since they transform as
\be
\psi_x  \ra (\Om_x P_L+\Om^*_x P_R) \psi_x   \;,\;\;\;
\psb_x \ra \psb_x (\Om_x  P_L+\Om^*_x P_R)   \;. \lb{EEET}
\ee
After inserting (\eq{PSPR}) into (\eq{YUKPR}) we obtain
an equivalent form of the action in terms of the $\psi$ and $\psb$ fields
%
% FIGURE 1
%
\begin{figure}[t]
\centerline{
\fpsysize=14.0cm
\fpsbox{Fig1.ps}
}
%\vspace*{12.0cm}
\vspace*{-1.0cm}
\caption{ \noindent {\em The $\kp$-$y$ phase diagram of the ISF model.
The solid lines represent the phase transitions between the various
phases. All phase transitions appear to be of second order.
}}
\label{FIG1}
\end{figure}
\bea
S_{U,V,\psi}\!\!\!&=&\!\!\!-\frac{1}{2} \sum_{x \mu}\left\{
\psb_x \gm_{\mu} (U_{\mu x} P_L+ U_{\mu x}^* P_R)
\psi_{x+\hmu} - \psb_{x+\hmu} \gm_{\mu} (U_{\mu x}^* P_L+U_{\mu x} P_R)
\psi_x \right\} \nonumber \\
\!\!\!& &\!\!\!  - y \sum_x \psb_x (V_x^2 P_L+ {V_x^*}^2 P_R) \psi_x
\;.\lb{YUK}
\eea
The bare mass term in (\eq{NAIV}) has now turned into a Yukawa term with
Yukawa coupling $y$.
Since gauge invariance is broken in the model without $V$ field,
one
may add a mass counterterm for the gauge bosons, which takes the
form of a kinetic term for the
radially frozen scalar field in the gauge invariant
extension, cf. eq.~(\ref{YMH}) above,
\be
S_{U,V} =  \kp \sum_{\mu} \left\{ V_x^* U_{\mu x} V_{x+\hmu} +
V_{x+\hmu}^* U_{\mu x}^* V_x
\right\} \;.\lb{HIGGS}
\ee

Models like (\eq{YUK})
have been studied
in the global symmetry limit with gauge interactions turned off,
i.~e. at $g=0$ with $U_{\mu x}=1$, where they reduce to pure
Yukawa models.
The phase diagram of the staggered fermion version  (ISF) of the model is
displayed  in fig.~\ref{FIG1}.
The full details of this model and its phase
diagram will be explained in sects. 3 and 4.
Beside the ferromagnetic (FM) and antiferromagnetic
(AM) phases at large positive and negative $\kp$
there are two different symmetric phases in which the magnetization
$v=\lag \sum_x V_x \rag/volume$ vanishes.
These two phases in fig.~\ref{FIG1}, PMW and PMS,
are separated by a phase transition line
$y_c(\kp)$, of presumably second order.
The phase diagram
is invariant
under a $\kp \ra -\kp$ reflection which
interchanges the FM and AM broken phases, cf.~sect.~4.
The details of the phase diagram
depend on the charge of the scalar field in the Yukawa
coupling: instead of the charge two field $V^2$ we could use
for the discussion of the symmetry restoration also
the charge one field $V$. A
%more generic
phase diagram for fermion-Higgs models
where the $\kp \ra -\kp$ symmetry is absent and the fermions couple to
$V$ in stead of $V^2$ can be found in ref.~\cc{PHASE}.
There the two symmetric phases are separated by a
funnel-like region containing a ferromagnetic phase.
Such non-universal features are, however, not important for our purpose
here.

At small $y<y_c(\kp)$
the Yukawa model has three scaling regions:
(A) One can approach the FM(W)-PMW phase transition from
within the FM(W) phase. Then the spectrum contains
massive fermions and in addition there is a Higgs particle
and a massless Goldstone boson.
(B) If one approaches the FM(W)-PMW phase transition from within the
PMW phase the spectrum contains massless fermions, but also two light
scalar particles associated with the $V$ field.
(C) Anywhere in the PMW phase, away from the
phase boundaries, the scalar particles
decouple and even though $y>0$, the spectrum contains only free
massless fermions.
These properties agree qualitatively with treating the Yukawa
coupling term in the action (\ref{YUK}) as a perturbation. By
interpreting $U_{\mu x}$ in (\ref{YUK}) as an external gauge
field, we draw the important conclusion that the fermions at small
$y$ are charged.

%%%%%%%
{}From the phase diagram of the Yukawa theory at small $y$ we infer
corresponding scaling regions (A-C) in the full theory with
dynamical gauge fields, at small gauge coupling $g$. In (A) we
have the Higgs phenomenon and massive fermions, in (B) we have
electrodynamics with massless fermions and in addition massive
charged scalars, while in (C) the scalars decouple. In the
region (C) the desired gauge symmetry restoration takes place
and we recover the target model: massless fermions
interacting with the U(1) gauge fields.

For large $y$ ($y > y_c(\kp)$) we do not recover the physics of
the target model.
Previous investigations have shown that
the effective action which describes the low energy physics in the PMS phase
is
more like eq.~(\eq{YUKPR}) and not like (\ref{YUK}).
The fermion spectrum contains only the
$\psi^\pr$ fermions whose couplings to the bosonic particles vanish as
a power of the lattice distance \cc{PMS}.
The $\psi^\pr$ fermions are neutral under the U(1) gauge symmetry.
They have masses, even in the PMS phase, which are generically
of the order of the cut-off.
Depending on details of the lattice model,
the mass in the PMS phase can be tuned to zero
(as in the
Wilson-Yukawa models for a sufficiently large value of the
Wilson-Yukawa coupling).
In any case the scaling physics in
the PMS phase is not that of the gauge invariant target model
(\eq{AXCO}) at $y=0$
because the fermions are neutral.

In section 4 we shall consider yet another way of breaking
the gauge invariance in the lattice models, namely, by changing
the normalization of $\gm_5$, making the replacement
\be
\gm_5 \ra \kp_A \gm_5,
\ee
with $\kp_A\neq 1$. This breaks gauge invariance on the lattice,
which is compact U(1), although in the classical continuum limit
the model is still invariant under non-compact gauge
transformations. Nevertheless, we shall show that in the quantum
case there is a wide range of values of $\kp_A$ for which there
is a PMW phase where symmetry restoration will take place.

Summarizing, our arguments for symmetry restoration in the
chiral gauge-fermion models with staggered  fermions
depend crucially on the existence of
a PMW phase in the global symmetry limit. This phase has the massless
fermions of the classical target model, which interact gauge
invariantly with the gauge field when it is switched on as an
external field. In the interior of the PMW phase (region (C))
there are no scaling scalars. Hence, for dynamical gauge fields
this may be considered as a satisfactory regularization of the
target model.

Strictly speaking the models studied here numerically are not
chiral since the charges $q_f$ are chosen, for numerical reasons,
such that the models are equivalent to vector models. However,
the existence of a PMW phase depends presumably only on the relative
smallness of the symmetry breaking and not on the chiral
properties of the theory.
\section{Staggered fermion models with axial-vector couplings}
In this section we shall introduce lattice versions of the continuum
model (\eq{AXCO}) which allows for a chiral set of charges
$q_f$, using staggered fermions. This is most easily done
in terms of $4\times 4$ matrix fields
$\Psi^{\al \kp}_x$ and $\Psb^{\kp\al}_x$, where the indices $\al$
and $\kp$ act as Dirac and flavor indices, respectively \cc{Sm88}.
Consider the following ansatz for a lattice action,
\bea
 S_{U,\Psi^\pr} \!\!\!&=&\!\!\!  -\frac{1}{2}\sum_{x\mu} \Tr \left\{
  \Psb_x^\pr \gm_{\mu}(U^{LL}_{\mu x} P_L
  + U^{RL}_{\mu x} P_R)\Psi_{x+\hmu}^\pr P_L
  -  \Psb_{x+\hmu}^\pr \gm_{\mu}({U^{LL}_{\mu x}}^*  P_L
+ {U^{RL}_{\mu x}}^*  P_R)\Psi_x^\pr P_L \right. \nonumber \\
 \!\!\!& &\!\!\! \left.  + \Psb_x^\pr \gm_{\mu}(U^{LR}_{\mu x}P_L
+  U^{RR}_{\mu x} P_R)\Psi_{x+\hmu}^\pr P_R
-  \Psb_{x+\hmu}^\pr \gm_{\mu}({U^{LR}_{\mu x}}^*  P_L
+ {U^{RR}_{\mu x}}^*  P_R)\Psi_x^\pr P_R \right\} \nonumber \\
\!\!\!& &\!\!\! - y \sum_x \Tr \left\{ \Psb_x^\pr \Psi_x^\pr \right\}
\;.\label{MODPSI}
\eea
The projectors $P_{L,R}$ on the left-(right-)hand side of $\Psi^\pr$
project on eigenstates of $\gm_5$ in the Dirac (flavor) space.
The $U_{\mu}^{LL},\ldots,U_{\mu}^{RR}$ may carry different
representations of the gauge group. We restrict ourselves here to
$\gm_5$ as the only non-trivial flavor matrix.

As it stands, the action (\eq{MODPSI}) has the same local gauge invariance
as the $N_F=4$
classical continuum model (\eq{AXCO}) if all
components $\Psi^{\al \kp}_x$ and $\Psb^{\kp \al}_x$ would be
independent degrees of freedom. However, such a model would have fermion
doublers and would lead to a vector-like gauge model for all choices of
$U^{LL},\ldots,U^{RR}$.
The fermion doublers are situated at the boundary of the
Brillouin zone in momentum space, i.e. near momenta $p_{\mu} =\pm
\pi$. After restricting the momenta of the matrix fields such
that
\be
-\pi/2<p_{\mu} \leq  +\pi/2\;, \lb{REG}
\ee
we loose the fermion
doublers of the matrix field, but also gauge invariance.
It is clear, however, that in the classical continuum limit, where the
fields and their momenta go to zero (in lattice units),
eq.~(\eq{MODPSI}) with the momentum restriction  (\eq{REG})
goes over into the corresponding continuum model.
Hence the full gauge invariance gets restored in this limit.

One might think that the cut-off in momentum space has to result in a
non-local action. However,
it is possible to express the action in a form that
is local, using staggered fermion fields $\chi_x$ and $\chb_x$.
The connection with the $\Psi^{\al\kp}_x$ and $\Psb^{\kp\al}_x$ is made
by writing \cc{Sm88,Sm92}
\be
\Psi_x=\eighth \sum_{b}  \gm^{x+b}       \chi_{x+b} \;,\;\;\;
\Psb_x=\eighth \sum_{b} (\gm^{x+b})^{\dg}\chb_{x+b} \lb{SDT} \;,
\ee
where
$\gm^x \equiv \gm_1^{x_1} \gm_2^{x_2} \gm_3^{x_3}\gm_4^{x_4}$
and the sum extends over the 16 corners of a unit lattice
hypercube, $b_{\mu}=0,1$. The substitution of
(\eq{SDT}) into the action (\ref{MODPSI}) leads to a local action.
In momentum space we have the relation \cc{Sm92}
\be
\Psi_{\al\kp} (p) = Z(p) \sum_b T_{\al\kp, b} (p)\chi (p+\pi b),\;\;\;
 -\pi/2 < p_\mu \leq +\pi/2 \;, \lb{MSPACE}
\ee
where $T_{\al\kp,b} = \sum_c \exp(ibc\pi)\gm^c_{\al\kp}/8$
is a unitary matrix and $Z(p)$ is a
non-vanishing function in the restricted momentum interval.
This relation is derived in the appendix.
It shows clearly that in the restricted momentum interval the Fourier
components of the matrix field are independent.

By choosing appropriate link fields $U_{\mu}^{LL}, \ldots,U_{\mu}^{RR}$, the
fermions couple in different ways to the gauge field.
For example, with
\be
U_{\mu}^{LL} = U_{\mu}^{RL}=U_{\mu}^{LR}=U_{\mu}^{RR}\;, \lb{U1}
\ee
the four fermions couple vector-like; with
\be
U_{\mu}^{LL} = U_{\mu}^{LR}=U_{\mu}\;,\;\;\;\; U_{\mu}^{RL}=U_{\mu}^{RR}=1
\lb{U2}
\ee
we have a left model in which only the four left-handed fermions couple
to the gauge field;
with
\be
U_{\mu}^{LL} = U_{\mu}^{LR}=U_{\mu}\;,\;\;\;\;
U_{\mu}^{RL}=U_{\mu}^{RR}=U_{\mu}^* \lb{U3}
\ee
we have a four fermion version of the axial-vector model (\eq{AXCO}) with all
$q_f=+1$ and finally with
\be
U_{\mu}^{LL} = U_{\mu}^{RR}=U_{\mu} \;,\;\;\;\;
U_{\mu}^{LR}=U_{\mu}^{RL}=U_{\mu}^* \lb{U4}
\ee
we have the axial-vector model (\eq{AXCO}) with $q_{1,2}=+1$ and $q_{2,3}=-1$.

The key issue of gauge symmetry restoration can be
investigated in model (\eq{MODPSI}, \eq{U3}).
It is advantageous to choose this axial vector model, and not the left model
(\eq{MODPSI}, \eq{U2}), because  the staggered version of this
model has a larger lattice symmetry group which reduces the
number of counterterms in the scaling region. Generalizing
$U_{\mu} \ra U_{\mu}^q$ in (\ref{U3}) to actions with arbitrary
integer charge $q$, it is straightforward to construct chiral
anomaly free models by combining several actions
(\ref{MODPSI},\ref{U3}), e.g. $5(q=+1) + 4(q=-2) + 1(q=+3)$. For
numerical reasons we shall however study only the simple $1(q=+1) +
1(q=-1)$ anomaly free model,
which is equivalent to a vector theory (eight flavor QED, for $y=0$),
obtained by adding mirror fermions (cf. sect. 4).
Model (\ref{U4}) with an equal number of $+1$ and $-1$ axial charges
can be easily modified such that it is exactly gauge invariant
for $y=0$ (see below). This gauge invariant version is the earlier
mentioned ISF model which is a
reduction of the naive fermion model in eq.~(\eq{NAIV}) and which serves here
as
a reference model.

Similarly, one can construct staggered fermion models involving arbitrary
coupling of the  Dirac-flavor components of the staggered flavors
to gauge or Higgs fields in a straightforward manner
such that the target models are recovered in the classical
continuum limit. We have shown that this works satisfactory in the
two-dimensional U(1) version of the equal charge axial model
\cc{SMOOTH}. For the Standard Model and Grand Unified
Theories like SO(10) and SU(5) and the axial-vector model
one can write down models in which the
staggered fermion symmetry group is preserved \cc{Sm92}.
This strategy of
coupling the Dirac-flavor components of the staggered flavors
has also recently been successfully
applied to a fermion-Higgs model \cc{UP}.

After the substitution (\eq{SDT}) and working out the trace in (\ref{MODPSI})
one obtains the action in terms of the staggered fermion fields.
For the equal charge axial model (\eq{MODPSI}, \eq{U3}), we find
\bea
 && S_{U,\cp} = - \frac{1}{2} \sum_{x\mu}\left[ c_{\mu x}
 \frac{1}{16}\sum_b \et_{\mu x+b}
    (\cpb_{x+b} \cp_{x+b+\hmu} - \cpb_{x+b+\hmu} \cp_{x+b})
                 %  \phantom{\frac{1}{2}}
 \right. \nonumber\\
 & &  +  \left. i s_{\mu x} \frac{1}{16}\sum_{b+c=n}
 \et_{5 x+b} (\et_{\mu x+c}
      \cpb_{x+b} \cp_{x+c+\hmu} - \et_{\mu x+b} \cpb_{x+b+\hmu} \cp_{x+c})
     \right] -y \sum_x \cpb_x \cp_x \;,  \label{NISF}
\eea
where
\be
c_{\mu x}=\RE U_{\mu x}\;,\;\;\;\; s_{\mu x}=\IM U_{\mu x}\;,\;\;\;\;
n=(1,1,1,1)\;. \lb{SC}
\ee
The staggered sign factors $\eta_{\mu x}$ are defined as
$\eta_{\mu x}=(-1)^{x_1+ \ldots + x_{\mu-1}}$.
The factor $\et_{5 x}=-\et_{4 x} \;\et_{3 x+\hat{4}}
\;\et_{2 x+\hat{3}+\hat{4}}\;\et_{1 x+\hat{2}+\hat{3}+\hat{4}}
= -(-1)^{x_1+x_3}$ represents the Dirac $\gm_5$.
In the classical continuum limit this action describes $N_F=4$
flavors of axially coupled Dirac fermions with $q_f=+1$,
$f=1,\ldots,4$.

For the axial model (\eq{MODPSI}, \eq{U4}) with
$q_{1,2}=+1$ and $q_{3,4}=-1$,
we find similarly,
\bea
   S_{U,\cp} \!\!\!&=&\!\!\! - \frac{1}{2} \sum_{x\mu}
\left[   c_{\mu x}
   \et_{\mu x}
      (\cpb_{x} \cp_{x+\hmu} - \cpb_{x+\hmu} \cp_{x})
   \right. \nonumber\\
   \!\!\!& &\!\!\!  +  \left. i s_{\mu x} \ep_{x} \et_{\mu x}
        (\cpb_x \cp_{x+\hmu} - \cpb_{x+\hmu} \cp_x)
       \right] - y \sum_x \cpb_x \cp_x \;.  \label{ISF}
\eea
%%%%%%%%%%%%%%%%%%%
The sign factor $\ep_x=(-1)^{x_1+ \ldots+x_4}$ represents the product of
the two $\gm_5$'s in Dirac and flavor space.%%%%
\hspace{0.1cm}From the substitution (\eq{SDT}) we would find
hypercubical
averages as in (\eq{NISF}) above, but we have chosen to replace those
according to $ \frac{1}{16} \sum_b f_{x-b} \ra f_x$. This leads
to the same classical continuum limit and makes the model
invariant under local gauge transformations (see below).
The form (\ref{ISF}) also follows by applying the familiar `spin
diagonalization' transformation  \cc{KaSm81}
to the naive
fermion action (\ref{NAIV}), which produces a sum of four actions
equivalent to (\ref{ISF}), and keeping only one of these.

In the equal charge axial model (\eq{NISF}) the projection on eigenstates
of the Dirac $\gm_5$ leads to non-local couplings in the $\h_5$ term
of (\eq{NISF}), which prevents
the gauge invariance of the naive fermion model (\ref{MODPSI})
to be carried over to
its staggered fermion
reduction.
Hence we shall
refer to this model as the non-invariant staggered fermion (NISF) model.
In a similar fashion we would loose gauge invariance in the left model
(\eq{MODPSI}, \eq{U2}) with staggered fermions.

In the model (\eq{MODPSI}, \eq{U4}) however,
the projection involves an additional $\gm_5$ in flavor space and
the model is invariant under the local `$\ep$-symmetry' of the staggered
fermions: the  action (\ref{ISF}) with $y=0$ is invariant
under the local flavor non-singlet U(1) transformation,
\be
 \cp_x   \ra  \exp (i \om_x \ep_x) \cp_x \;,\;\;\;
 \cpb_x  \ra  \cpb_x \exp (i \om_x \ep_x )  \;, \;\;\;
 U_{\mu x}  \ra  \exp (-i \om_x) U_{\mu x} \exp (i \om_{x+\hmu}) \;.
   \lb{EPSYM}
\ee
Therefore we shall call the model (\eq{ISF})
in the following the invariant staggered fermion (ISF) model.
As delineated in sect.~2, gauge symmetry is restored in this
model also for $y>0$, provided
$y$ is smaller than $y_c$. By making the
transformation of variables,
$\cp_x =  \exp (i \om_x \ep_x) \chi_x$,
$\cpb_x =  \chb_x \exp (i \om_x \ep_x )$ with $V=\exp i \om_x$ we find
the Yukawa form of the action in terms of the $\chi$ and $\chb$ fields
\bea
 S_{U,\chi} \!\!\!&=&\!\!\! - \frac{1}{2} \sum_{x\mu}
\left[   c_{\mu x}
   \et_{\mu x}
      (\chb_{x} \chi_{x+\hmu} - \chb_{x+\hmu} \chi_{x})
   \right. \nonumber\\
   \!\!\!& &\!\!\!  +  \left. i s_{\mu x} \ep_{x} \et_{\mu x}
        (\chb_x \chi_{x+\hmu} - \chb_{x+\hmu} \chi_x)
       \right]
   - y \sum_x \left[\RE V_x^2 +i \ep_x \IM V_x^2 \right] \chb_x
\chi_x \;,
                                   \label{ISFYM}
\eea
with the U(1) charge two
scalar field $V^2$. We remark that now, in contrast
to the naive model, the fermion measure $D\chb D\chi$ is not invariant
under the transformations
(\eq{EPSYM}). The invariance of the measure is however recovered,
after adding mirror fermion fields, which is required also
for the use of the Hybrid Monte Carlo Algorithm (HMCA)
in the numerical simulations,
cf.~sect.~4. So the final ISF model has four $q_f=+1$ and four
$q_f=-1$.

In contrast to the ISF model the NISF model has lost its gauge invariance.
The model can be thought of as being (almost) gauge
invariant for the low momentum components of the gauge field
while the symmetry violation increases for the high momentum
components.

Following the procedure explained in the previous section, the
gauge degrees of freedom in the non-invariant model can be
interpreted as scalar fields and the action with those scalar fields
is obtained by replacing the sine and cosine factor in (\eq{SC}) by
\be
  c_{\mu x} = \RE V_x^* U_{\mu x} V_{x+\mh}\;,\;\;\;\;
  s_{\mu x} = \IM V_x^* U_{\mu x} V_{x+\mh}\;.   \lb{CSVUV}
\ee
In our numerical work we shall furthermore restrict ourselves to the
zero gauge coupling limit $U_{\mu}=1$.

We have also introduced a further modification of the models that
provides an additional parameter to monitor a possible restoration of
gauge invariance. As can be seen in the staggered action (\eq{NISF}), the
$\gm_{\mu}$ part of the target action leads only to a one-link
coupling among the $\cp$ and $\cpb$ fields, whereas the
$\gm_{\mu} \gm_5$ part involves three-and also five-link couplings.
Therefore we expect that the $\gm_\mu \gm_5$ part of the action
renormalizes differently from the $\gm_\mu$ part and we may
have to compensate
for this with a finite renormalization factor in front of the $\gm_\mu\gm_5$
term.
This leads to the  replacement
\be
s_{\mu x} \ra \kp_A s_{\mu x}  \label{KAPPAA}
\ee
in the action (\eq{NISF}). We can also modify the ISF model in this
way. For $\kp_A \ne 1$ the gauge invariance is
broken and this model may be more similar to the NISF model.

For simplicity we use the same $\kp_A$ for the three and five
link couplings.
The appearance of both three and five link couplings in the
$\gm_\mu \gm_5$ term, is an
awkward feature of the action
(\eq{NISF}), because the three and five link terms may
renormalize differently. Furthermore, a canonical
construction of a transfer
matrix  requires that the couplings are confined within a hypercube.
We can arrange for this, by replacing
the five link terms in (\eq{NISF}) by equivalent three link
terms, which have the same classical continuum limit.
We have sometimes used
such a modified form of the
action instead of (\eq{NISF}), but did not find a
significant difference in the measured observables.

The crucial question now is whether the breaking of gauge
invariance in the NISF model is sufficiently weak that we can
have symmetry restoration in a PMW scaling region. The mechanism
for this would be similar to that in the ISF model, for which the
gauge degrees of freedom $V_x$ can be explicitly transformed into
a Yukawa term (cf. (\ref{ISFYM})), leading to charged
massless fermions in the symmetric phase at weak Yukawa coupling.
For the NISF model we cannot explicitly transform the $V_x$'s
into a Yukawa coupling, but the same mechanism may take place
effectively in the low momentum modes where the symmetry breaking
is weak. The high momentum modes might primarily lead to a
`renormalization of $\gm_5$', to be compensated by $\kp_A$. We
can get an idea of the sensitivity to a mismatch in $\kp_A$
values by studying the effect of $\kp_A\neq 1$ in the ISF model.
A major danger to the above mentioned scenario is that unlike the
case of the massive Yang-Mills model and the ISF model, we cannot
easily control the strength of the gauge symmetry breaking in
the NISF model.
\section{Phase diagram of the ISF model}
To determine the phase structure of the ISF and NISF models we
shall compute numerically the dependence of a number of local order
parameters on the bare coupling parameters $\kp$, $y$ and $\kp_A$.
We use the Hybrid Monte Carlo method, which requires a positive definite
fermion determinant. The fermion matrix for the ISF and NISF models can
be obtained from the actions (\eq{ISF}) and (\eq{NISF}) by writing them
in the form
\be
S_{ISF} = -\cpb M_{ISF}\cp \;,\;\;\;
S_{NISF} = - \cpb M_{NISF}\cp \;.          \lb{DEFM}
\ee
The fermion determinant,
$\Det M$, is not positive definite, but
this we cure by replacing it by
$\Det(M^\dg M)$ which is
manifestly positive definite. This amounts to adding a mirror fermion
field to the action, which couples to the matrix
$M_{(N)ISF}^\dg$.
With this extra field both the ISF and the NISF model have equal number
of fermions coupling with charge $q_f=+1$ and $q_f=-1$, which makes them
vector-like and anomaly free.

We should mention
here that
the actions (\eq{ISF}) and (\eq{NISF}) for the ISF and NISF model
are invariant under the discrete $\ep$-symmetry
\be
\cp_x \ra \ep_x \cp_x\;, \;\;\cpb_x \ra  \cpb_x \ep_x \;, \;\;
V_x \ra \ep_x V_x  \;,\;\;  \kp \ra -\kp \;, \lb{EPS}
\ee
which implies that the $\kp$-$y$ phase diagrams of the two models
have to be invariant for $\kp \ra -\kp$. Furthermore
the actions are invariant under the symmetry
\be
\cp_x \ra i\ep_x \cp_x\;, \;\;\cpb_x \ra  \cpb_x i\ep_x \;, \;\;
y \ra -y \;, \lb{YAXIS}
\ee
which implies that the two phase diagrams are also symmetric around
the $y=0$ axis. It is therefore sufficient to restrict ourselves
to the half plane $y>0$.

The numerical simulations
were carried out mainly on a $4^4$ lattice. A few simulations were
also performed on larger lattices ($6^4$ and $8^4$).
We used
periodic boundary conditions for the scalar fields. For
the fermion fields periodic boundary conditions were used in
the spatial direction and antiperiodic boundary conditions in the time
direction.  We typically accumulated a statistics of 500-3000 HMCA
trajectories at each point in the phase diagrams, depending on the
autocorrelation time for the various observables.
The step size in the HMCA was adjusted such that the acceptance rate
was between 60 and 80\%.

The local observables which we shall consider are:  \\
\noindent the magnetization,
\be
v=
\left|\left\lag  \frac{1}{V} \sum_x V_x|_{\mbox{rot}}  \right\rag
\right| \;,
\ee
where $V$ is the lattice volume and
the subscript ``rot'' indicates that the standard rotation technique has
been applied to account for the drift of the magnetization vector
$(1/V) \sum_x V_x$ on a finite system \cc{ROT}; \\
\noindent the staggered magnetization,
\be
v_{st}=\left|\left\lag  \frac{1}{V} \sum_x \ep_x V_x|_{\mbox{rot}}
\right\rag \right|\; ;
\ee
\noindent the energy associated with the kinetic term
(\eq{HIGGS}) for the scalar field,
\be
E_{\kp}=
\left\lag \frac{1}{4V} \sum_{\mu x} \mbox{Re} (V_x^* V_{x+\hmu}) \right\rag
= \frac{1}{8V}
\frac{\del}{\del \kp} \ln Z \;,
\ee
with $Z$ the partition function; and \\
\noindent the energy associated with the mass term
for the fermions
\be
E_{y}= \left\lag \frac{1}{V} \sum_x \cp_x \cpb_x \right\rag
= \frac{1}{V}
\frac{\del}{\del y} \ln Z \;.
\ee
It also proved to be useful to keep track of the number
$N_{CG}$ of Conjugate
Gradient Algorithm iterations,
needed for an inversion of the fermion matrix
to a given accuracy. This quantity is large,
if the fermion matrix has eigenvalues close to zero.
The observables $v$ and $v_{st}$ are order parameters for
ferromagnetism and antiferromagnetism and are non-zero
in the FM and AM phases, respectively.
$\;\;$From the $\kp$-dependence of $v$ and $v_{st}$ we determined the
position of FM-PM and PM-AM phase transitions shown in fig.~\ref{FIG1}.
A zero fermion mass would be a clear signal of the PMW phase, but it
would require larger lattices to measure this mass reliably.
Instead we have read off the
position of the PMW-PMS phase transition
from the $y$-dependence of the
energy $E_y$.
The energies $E_{\kp}$ and $E_y$
can be used to get information about the order of the phase transition.

%
% FIGURE 2
%
\begin{figure}%[t]
 \centerline {
 \fpsysize=14.0cm
 \fpsbox{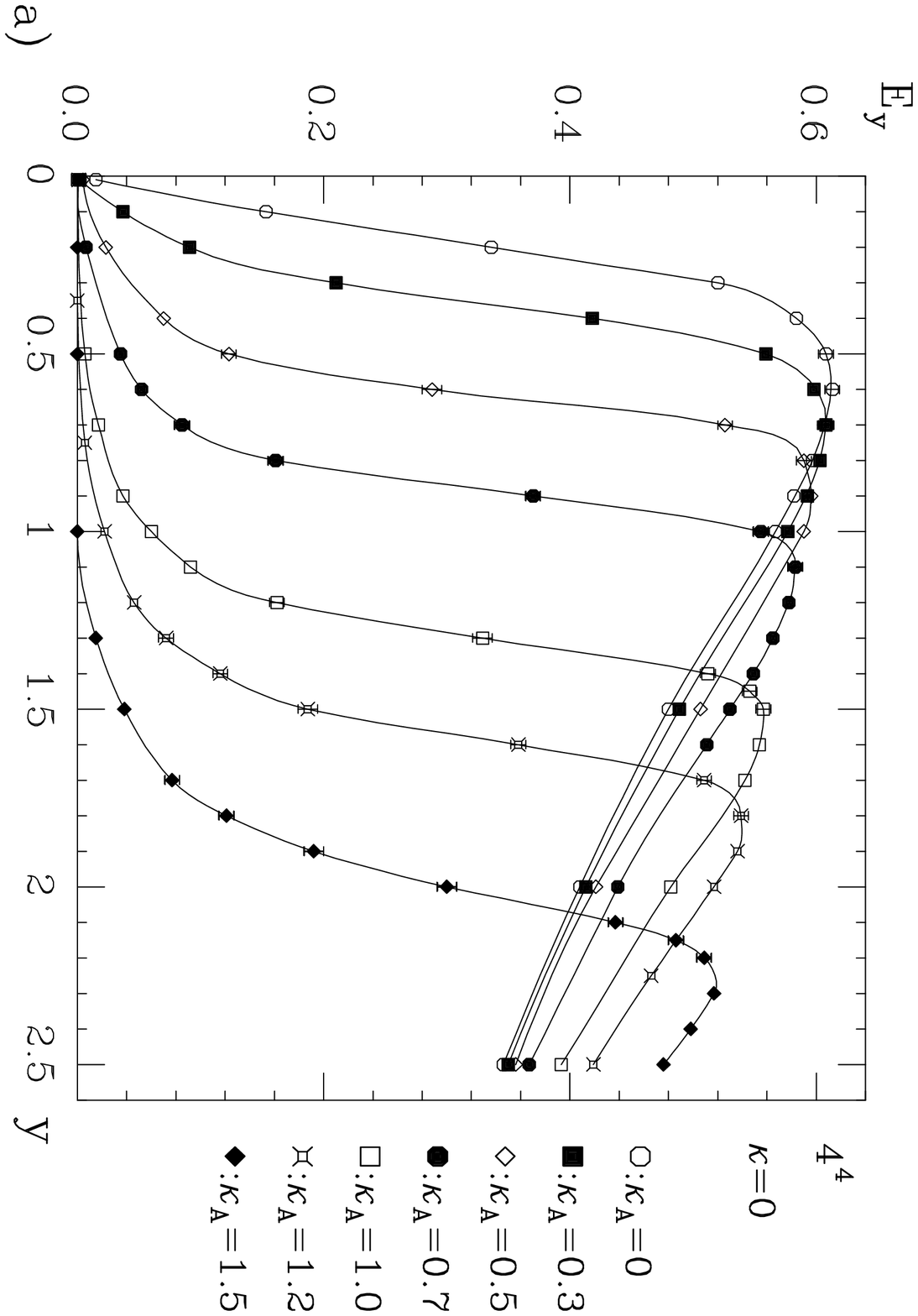}
 }
 \vspace*{-1.0cm}
 \centerline {
 \fpsysize=14.0cm
 \fpsbox{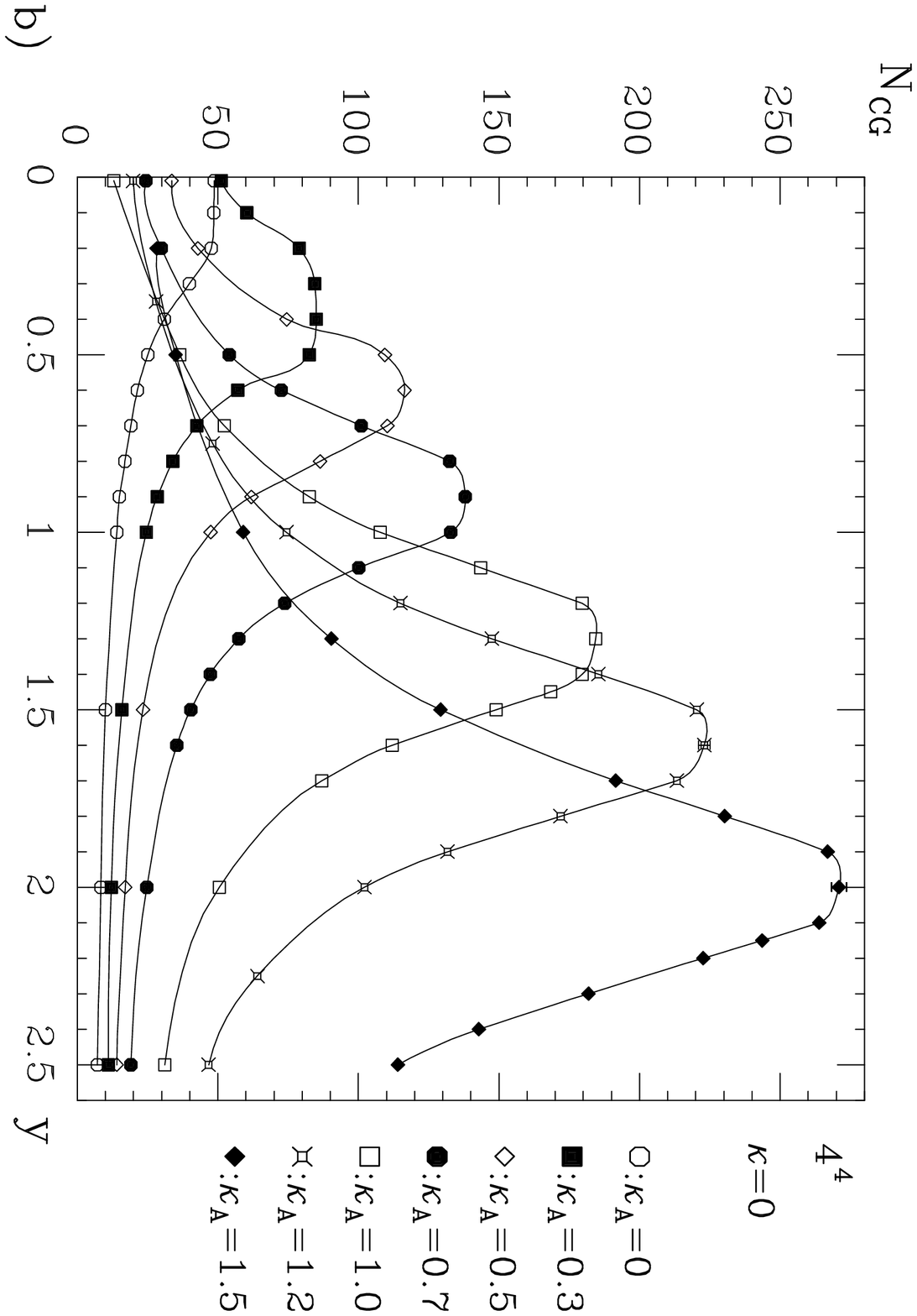}
 }
\vspace*{-1.0cm}
\caption{ \noindent {\em a)
The energy $E_y$ as a function of $y$ for several values of $\kp_A$ and
$\kp=0$ in the ISF model. b) The quantity $N_{CG}$ for the same set
of bare coupling parameters. Here and in the following we omit
the error bars whenever they are smaller than the symbol size.
}}
\label{FIG2}
\end{figure}
We now discuss the numerical results for the ISF model
in more detail.
In fig.~\ref{FIG2}a we have plotted the energy
$E_y$ as a function of $y$ for $\kp=0$ and several values of the $\gm_5$
renormalization factor $\kp_A$ (cf. (\ref{KAPPAA})).
The open squares were obtained
at  $\kp_A=1$, in which case the model is equivalent to the Yukawa
model (\eq{ISFYM}). The energy $E_y$ has its steepest
slope
at the PMW-PMS phase transition which
for $\kp_A=1$ is close to
$y=1.4$. At this point the curvature changes also its sign.
It appears that the phase transition is of
second order, which we infer from the absence of a jump in $E_y$
on larger lattices and the absence of a hysteresis
effect in thermal cycles.  The energy $E_y$ is very
small, but still non-zero in the PMW phase,
increases at
the PMW-PMS phase transition and is large in the PMS phase.
The approximate fall off $\propto 1/y$ at large $y$ can be
understood from the eigenvalue distribution of the fermion matrix
$M_{ISF}$: We can write $E_y=(1/V) \lag \mbox{Tr} M_{ISF}^{-1} \rag
= (1/V) \lag \sum 1/(\lm(y)+y) \rag$. The sum extends over all
eigenvalues $\lm$ of the off-diagonal part of the fermion matrix $M_{ISF}$
whose eigenvalue distribution turns out to be almost independent of $y$
(see below).

The squares in fig.~\ref{FIG2}b show the $y$-dependence of
$N_{CG}$.
This quantity exhibits a maximum at the critical value of $y$
indicating that the PMW-PMS phase transition is associated with
the occurrence of small eigenvalues of the fermion matrix $M_{ISF}$.
We will come back to this point at the end of this section.

%
% FIGURE 3
%
\begin{figure}[t]
 \centerline {
 \fpsysize=14.cm
 \fpsbox{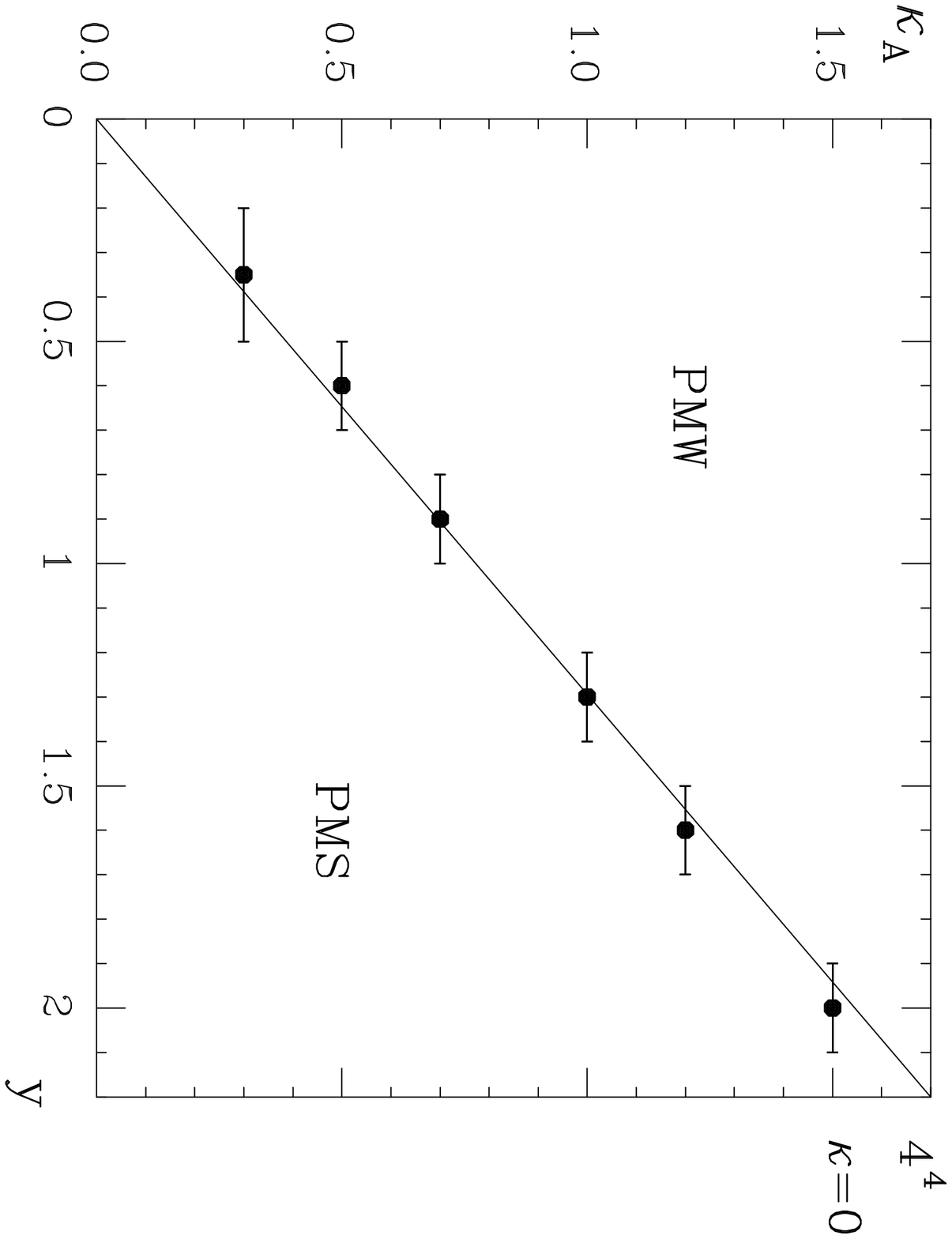}
 }
% \vspace*{12.0cm}
\vspace*{-1.0cm}
\caption{ \noindent {\em The
phase diagram of the IFS model in the $\kp_A$-$y$ plane for $\kp=0$.
The lattice size is $4^4$.
}}
\label{FIG3}
\end{figure}
As we mentioned in the previous sections, it is interesting to investigate
the ISF model with different $\kp_A$.
Gauge invariance is broken for $\kp_A\neq 1$
and we loose  the equivalence of the models (\eq{ISF}) and (\eq{ISFYM}).
In the figs.~\ref{FIG2}a and b we have also
included the results for $E_y$ and $N_{CG}$ for several values of
$\kp_A \ne 1$. Interestingly it can be seen from the shape
of the curves that a PMW-PMS phase transition emerges
for all $\kp_A > 0$. The position of
the PMW-PMS phase transition shifts to smaller values of $y$ when the
value of $\kp_A$ is lowered as is obvious from the $y$-dependence of the
points with the steepest slope in fig.~\ref{FIG2}a
and of the maximum in fig.~\ref{FIG2}b.

In fig.~\ref{FIG3} we have
plotted the approximate position of the PMW-PMS phase transition, as
determined from figs.~\ref{FIG2}a and b
for $\kp=0$ into a $\kp_A$-$y$ phase diagram. The data points
fall within error bars on a straight line through the origin with slope
$\approx 0.77$.
We encounter here another example of restoration of gauge invariance:
The gauge symmetry is broken in the lattice action, but
we expect that the PMW phase leads to the same
scaling
physics for
all $\kp_A>0$.
Only at $\kp_A=0$, where the $\gm_{\mu} \gm_5$
part disappears completely from the action (\eq{ISF}),
the curve for $E_y$ in fig.~\ref{FIG2} shows no indication for
a change of curvature at small $y$ and
the PMS phase extends most probably down to $y=0$.

%
% FIGURE 4
%
\begin{figure}[t]
 \centerline {
 \fpsysize=14.0cm
 \fpsbox{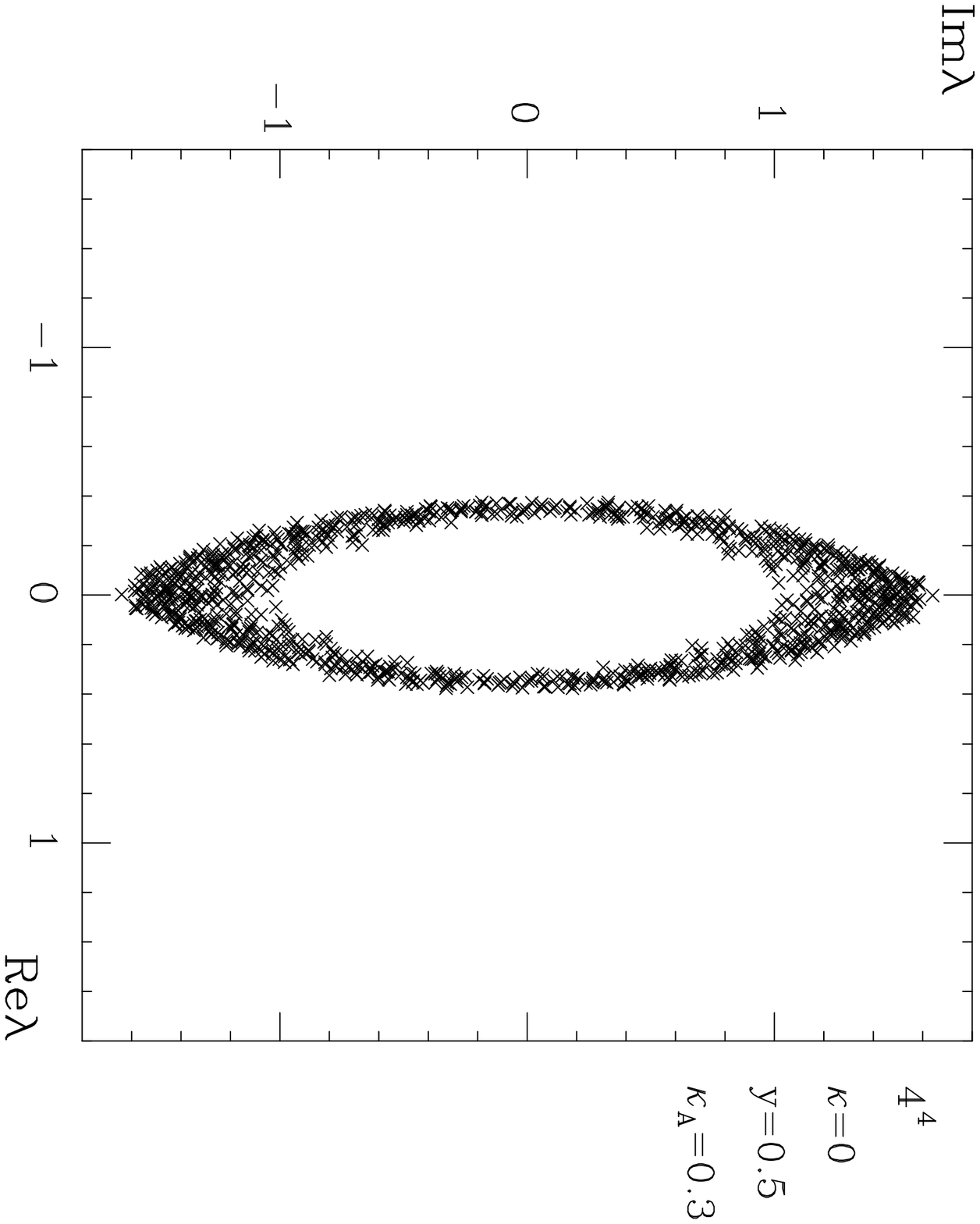}
 }
% \vspace*{12.0cm}
\vspace*{-1.0cm}
\caption{ \noindent {\em
The eigenvalues  of the matrix $M_{ISF}^{off}=M_{ISF}-y \un$ plotted
into the complex plane. The graph  contains the eigenvalue spectra for four
independent scalar field configurations which were generated with the
HMCA at $(\kp,y,\kp_A)=(0,0.5,0.3)$.
The lattice size is $4^4$.
}}
\label{FIG4}
\end{figure}
Some additional insight about the phase structure is obtained from
the eigenvalue spectra of the fermion matrix
$M_{ISF}$. The eigenvalues were computed with the Lanczos algorithm.
In fig.~\ref{FIG4} we have plotted the eigenvalues of the off-diagonal part
$M_{ISF}^{off}=M_{ISF}-y \un$
in the complex plane, for four independent scalar field
configurations which were generated with the HMCA on a $4^4$ lattice
at $\kp=0$, $y=0.5$ and $\kp_A=0.3$. We found empirically that the
eigenvalue distributions obtained at other values of
$y$, with $\kp_A$ and $\kp$ kept fixed, have almost identical shape.
In particular  they always cut the real axis at the same distance $h$
from the origin, which is $\approx 0.37$ for $\kp_A=0.3$.

The distribution of $M_{ISF}$ for some non-zero value of $y$
with $\kp_A$ kept fixed is then approximately obtained by shifting the
distribution shown in fig.~\ref{FIG4}
by an amount $y$ along the real axis. This explains the observed
behavior of $N_{CG}$ in fig.~\ref{FIG2}b. For $y<h$ the origin is
situated within the hole and $N_{CG}$ is small since there are no
eigenvalues near the origin. It develops a peak when $y \ra h$,
because eigenvalues come close to origin
and it decreases again when $y$ is increased beyond $h$. We conclude
that $y_c \approx h$, which is consistent with the value obtained
from fig.~\ref{FIG3}.

The width $h$ of the eigenvalue distribution increases for increasing $\kp_A$.
For $\kp_A=1$ we find the eigenvalues to be arranged along the boundary
of a circle and at larger $\kp_A$ along the boundary of an ellipse with
the principal axis on the real axis. At $\kp_A=0$ the fermion matrix is
anti-hermitian and the eigenvalues fall
exactly
on the imaginary axis.
This gives additional support for the
absence of a phase transition in $y$ at $\kp_A=0$.

As is illustrated with this discussion,
the appearance of a PMW phase at small $y$ is associated with a
hole in the eigenvalue distribution, which gives rise to the concave
$y$-dependence of $E_y$ for $y<y_c$ and to the peak in the number of CG
iterations at $y=y_c$. We shall also look for these features in the
NISF model to establish if a PMW phase emerges there, in which the
gauge symmetry breaking due to the high momentum scalar modes  is
restored.
\section{Phase diagram of the NISF model}
We have seen in the previous section that the existence of
the symmetry restoring PMW phase of the ISF model is stable
against large deviations of $\kp_A$ from its preferred value 1.
This suggests that a fine tuning of $\kp_A$ in the NISF model may
not be necessary and it should suffice to choose $\kp_A$
sufficiently large such that the PMW phase becomes clearly
visible.
In this section we discuss the phase diagram of the NISF model, mostly
at fixed $\kp=0$, but for various values of $\kp_A$.
(Preliminary results for
$\kp\neq 0$ were reported in the first reference in
\cc{SMOOTH}.)

\begin{figure}[t]
 \centerline {
 \fpsysize=14.0cm
 \fpsbox{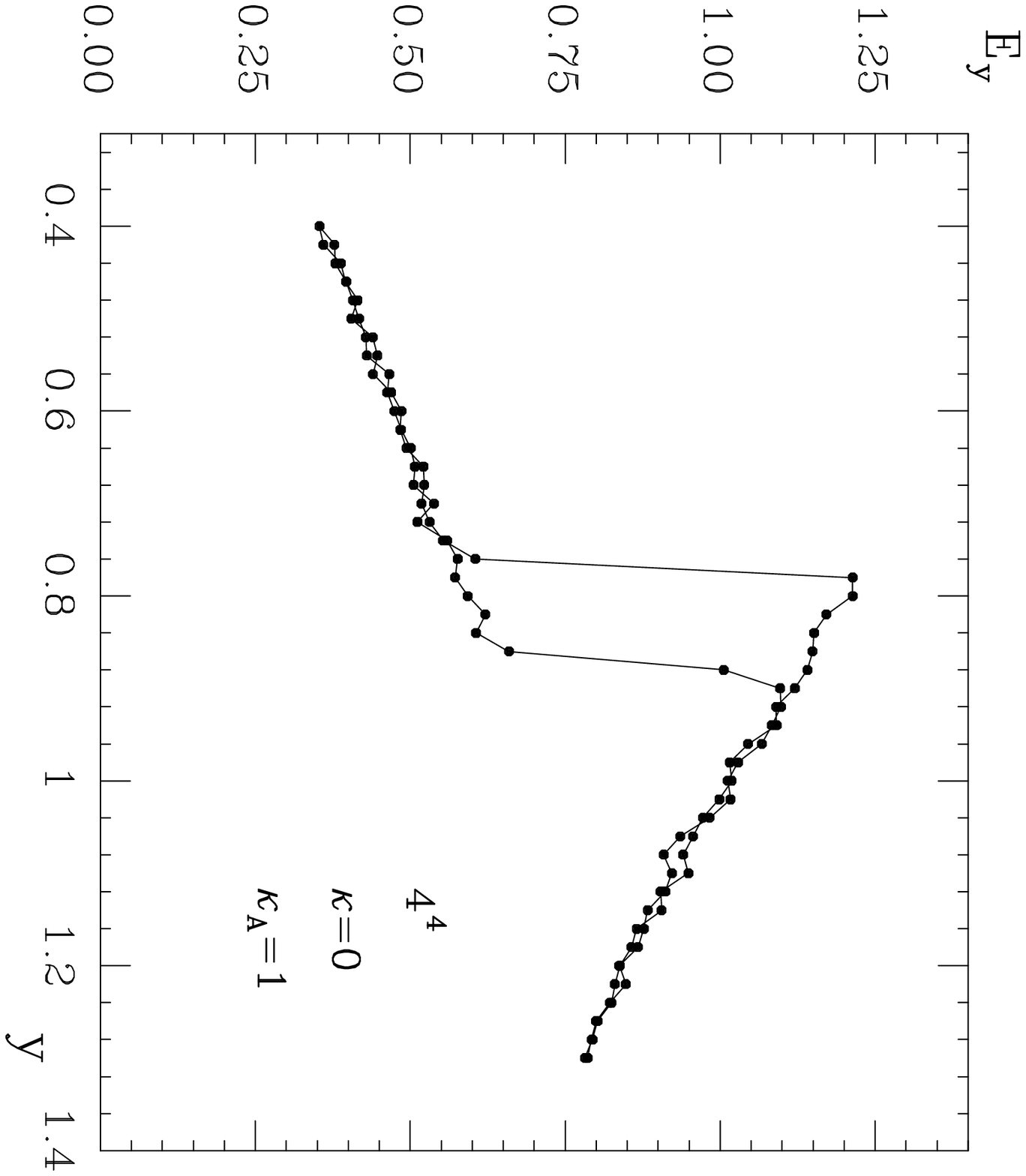}
 }
%\vspace*{12.0cm}
\vspace*{-1.0cm}
\caption{ \noindent {\em The energy $E_y$ as function of $y$ for
$\kp=0$ and $\kp_A=1$ in the NISF model.
The points were obtained in a thermal cycle starting at
$y=1.3$ with a step size of $\Delta y=0.02$. Each point was obtained
by taking the average over 50 HMCA trajectories.
}}
\label{FIG5}
\end{figure}

\begin{figure}[t]
 \centerline {
 \fpsysize=14.0cm
 \fpsbox{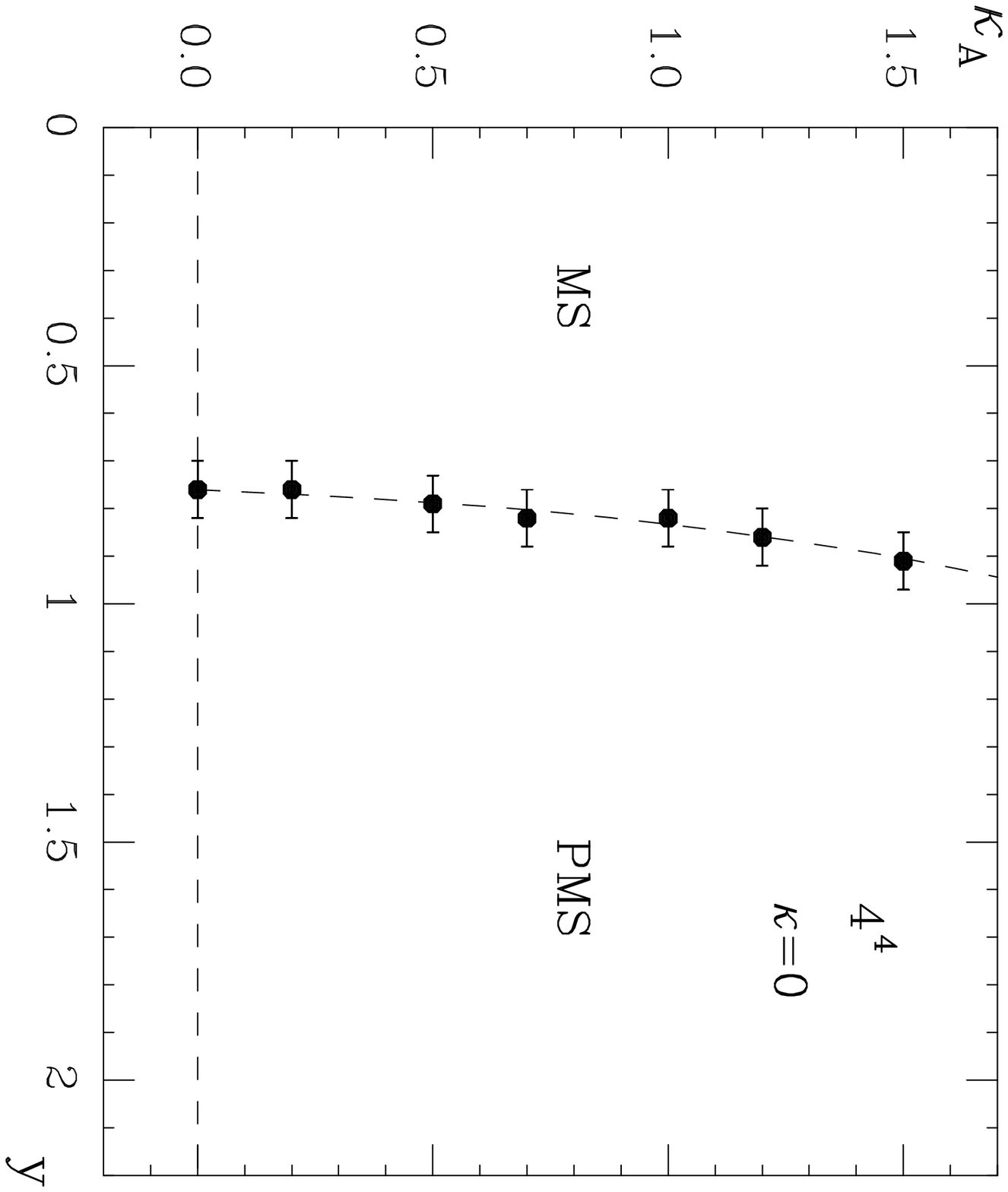}
 }
%\vspace*{12.0cm}
\vspace*{-1.0cm}
\caption{ \noindent {\em Phase diagram of the NISF model in the
$\kp_A$-$y$ plane obtained for $\kp=0$. The dashed
line indicates that the phase transition between the multi-state
(MS) region and the PMS phase is of first order.
}}
\label{FIG6}
\end{figure}

In fig.~\ref{FIG5} we have plotted the energy $E_y$ obtained in
a thermal cycle
as a function of $y$ for $\kp_A=1$. The behavior is quite different
from the one observed for this quantity in the ISF model. The energy
exhibits a jump at $y \approx 0.8$ and a clear hysteresis effect
is observed in the thermal cycle, indicating the presence  of a first
order phase transition. A similar behavior is found for
$\kp_A \ne 1$ and the position of the first order phase transition
in the $\kp_A$-$y$ phase diagram is
shown in fig.~\ref{FIG6}. Note that
the phase boundary does not go through the origin as in
fig.~\ref{FIG3} and that there is even a phase transition at
$\kp_A=0$.
%
% FIGURE 5
%

The properties of the
NISF model in the weak coupling region are unusual and quite different
from those in the PMW phase of the ISF model. From the  behavior
of the energy $E_{\kp}$ and the order parameters $v$ and $v_{st}$
we find that the system can go into one of the following
different states when lowering
the parameter $y$ from the PMS phase across the first order phase
transition: \\
\noindent FM$_{\mbox{ }}$: $v\neq 0$, $v_{st}=0$ and $E_{\kp}>0$. \\
\noindent PM$_+$: $v=0$, $v_{st}=0$ and $E_{\kp}>0$. \\
\noindent PM$_0$: $v=0$, $v_{st}=0$ and $E_{\kp}=0$. \\
\noindent PM$_-$: $v=0$, $v_{st}=0$ and $E_{\kp}<0$. \\
\noindent AM$_{\mbox{ }}$: $v=0$, $v_{st}\neq 0$ and $E_{\kp}<0$. \\
We shall therefore denote the weak coupling
region
in the following as a multi-state (MS)
region.
Also the number of CG inversions $N_{CG}$ shows a qualitatively different
behavior in the NISF model compared to the ISF model.
The quantity $N_{CG}$
grows when approaching the PMS-MS phase transition within the
PMS phase, then jumps to a small value when crossing this transition
and remains approximately constant within the MS
region.

%
% FIGURE 6
%
\begin{figure}[t]
 \centerline {
 \fpsysize=14.0cm
 \fpsbox{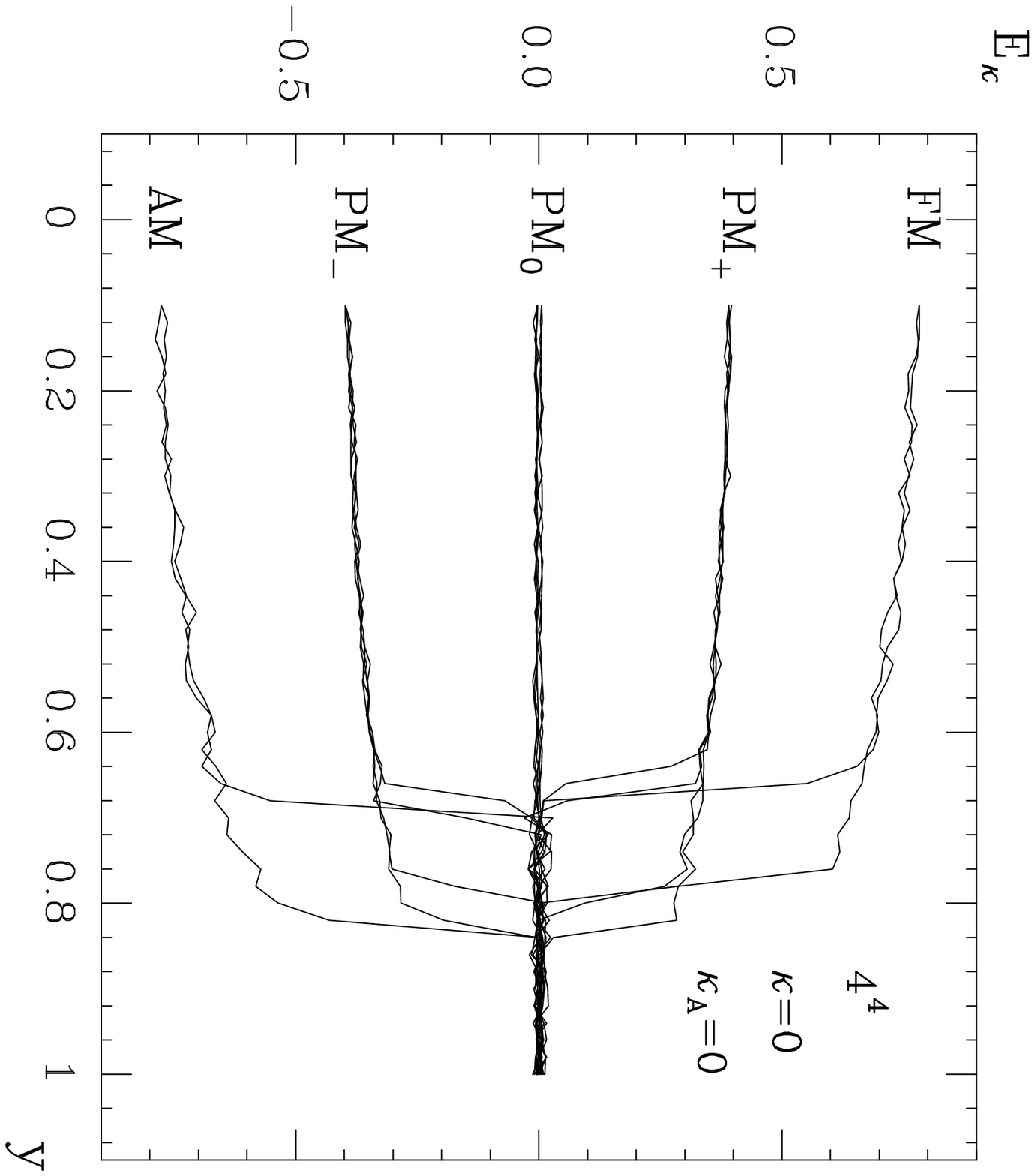}
 }
%\vspace*{12.0cm}
\vspace*{-1.0cm}
\caption{ \noindent {\em The energy $E_{\kp}$ as a function of $y$ for
$\kp=0$ and $\kp_A=0$ in the NISF model. The plot contains nine thermal cycles,
which were started at $y=1$ inside the PMS phase
with a step size $\Delta y=0.02$. The plot shows that
the system can tunnel into five different states when lowering $y$
across the MS-PMS phase transition.
Each point in a thermal cycle results
from averaging over 50 HMCA trajectories.
}}
\label{FIG7}
\end{figure}

Fig.~\ref{FIG7}
shows the energy $E_{\kp}$ as a function of $y$ at $\kp=0$ and $\kp_A=0$ for
several thermal cycles started at $y=1.0$ in the PMS phase.
The jumps in $E_{\kp}$ show
that the FM-PMS, PM$_+$-PMS, PM$_-$-PMS and
AM-PMS are of first order. Also the PM$_0$-PMS phase transition is
of first order since the energy $E_{y}$ exhibits a jump for all these
transitions. The appearance of five (meta) stable states in the PMW
phase indicates that the inclusion of the fermions produces additional
(local) minima in the effective action for the scalar field $V_x$.
The $\kp\ra -\kp$
symmetry (\ref{EPS}) relates the
PM$_{\pm}$ states and the FM with the AM state.
Even on very small lattice volumes the local minima appear to be
separated  by very high energy
barriers in  configuration space since we have never observed
tunneling events between the various states of the MS
region.
The frequency distribution of the states which
results after a repeated crossing of the PMS-MS transition
indicate that system has a slight preference for the PM$_0$ phase.

In contrast to the PMW phase of the ISF model, the MS region in the NISF
model persists for $0<y\aleq 0.7$ at $\kp_A=0$. This indicates that this
region is different from the PMW phase in the ISF model.
It is not excluded, however,
that a new phase, which harbors gauge symmetry restoration, is still
present for smaller values of $y$, hidden inside the MS region.
After scanning the various  states of the MS region in the $y$ and $\kp_A$
directions ($0 \leq \kp_A \leq 2$, $0.1 \leq y \leq y_c(\kp_A) $)
on an $8^4$ lattice we did not find any indication for the emergence of
another phase transition.

%
% FIGURE 7
%
At $\kp_A=0$ the $\gm_{\mu} \gm_5$ term in the fermion action is absent,
and the only difference between the
NISF model and the ISF model is the hypercubical average
over $\mbox{Re}(V_x^* V_{x+\hmu})$ in the one-link $\gm_\mu$ part of the
action. This implies that the phenomenon  of the MS region and the first
order phase transition is induced by this hypercubical
average. It suggests to study also a modified model with
the replacement
\be
\frac{1}{16} \sum_b \mbox{Re}(V_{x-b}^* V_{x-b+\hmu}) \ra
\mbox{Re}(V_{x}^* V_{x+\hmu})  \lb{SUB}
\ee
in the one-link term of the action (\eq{NISF}).
In the classical continuum limit this modified model leads
to the same target model.
The modification also
makes the NISF model more similar to the ISF model.

\section{Phase diagram of the modified NIFS model}
\begin{figure}%[t]
 \centerline  {
 \fpsysize=14.0cm
 \fpsbox{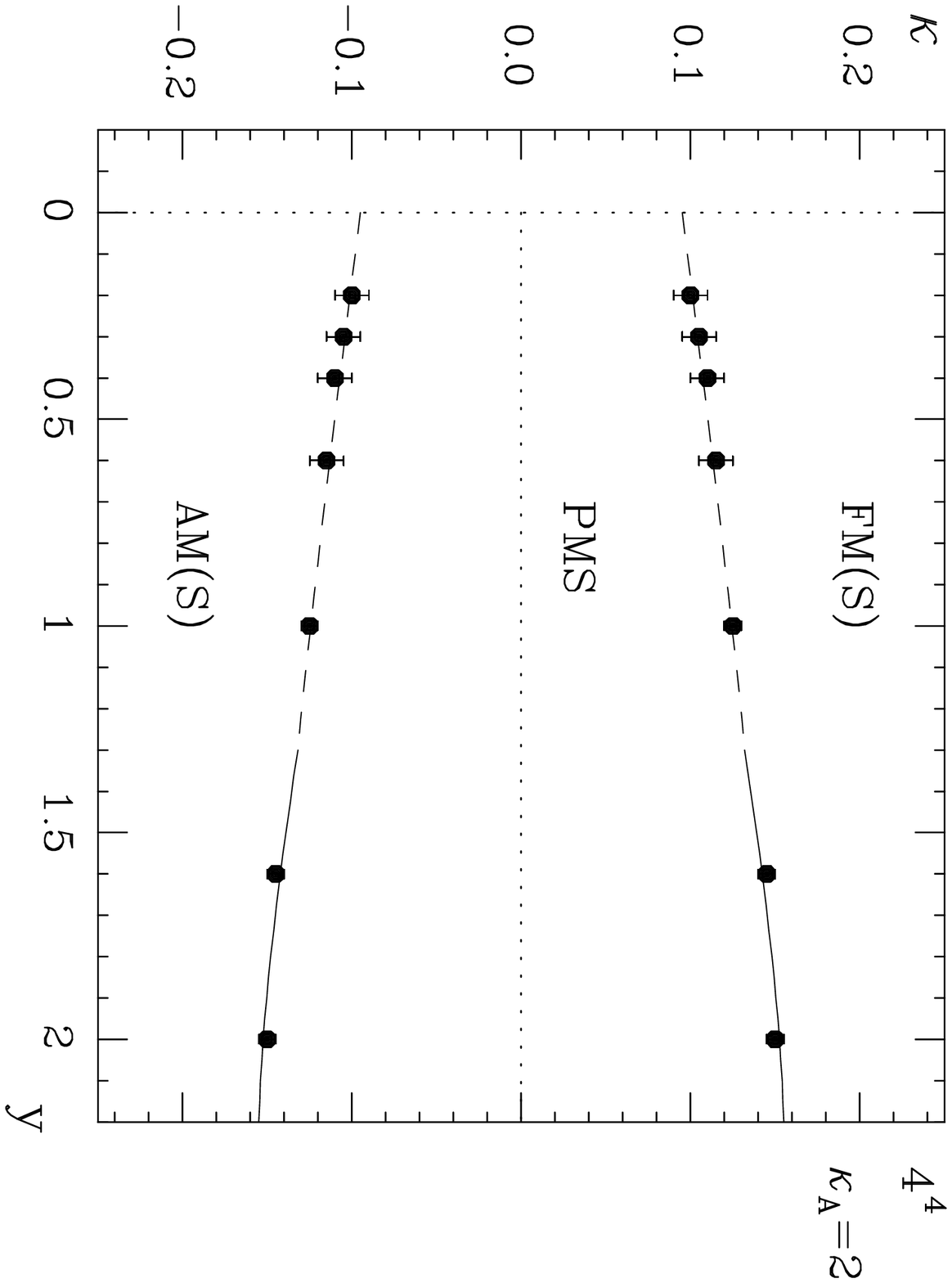}
 }
%\vspace*{12.0cm}
\vspace*{-1.0cm}
\caption{ \noindent {\em The $\kp$-$y$ phase diagram of the modified
NISF model for $\kp_A=2$.
The solid lines represent the phase transitions of second and the
dashed lines phase transitions of first order.
}}
\label{FIG9}
\end{figure}

{}From the behavior of the observables
$v$, $v_{st}$ and $E_{\kp}$ we obtain the phase diagram in the $\kp$-$y$
plane shown in fig. \ref{FIG9}, which applies to the case $\kp_A=2$. In
contrast
to the phase diagram of the NISF model without the replacement
(\eq{SUB}), we find no MS
region at small $y$.
The phase diagram contains only the strong coupling phases FM(S), PMS and
AM(S).  The
FM(S)-PMS and
PMS-AM(S) phase transitions are most likely of first order (dashed
line) for $y \lsim 1.4$ and of second order (full line) for $y \gsim 1.4$.
The gap in $E_{\kp}$, $v$ and $v_{st}$ is seen to increase
when $y$ is lowered.

%
% FIGURE 9
%
The $\kp$-$y$ phase diagram at smaller  values of $\kp_A$ looks very similar
to the one shown in fig.~\ref{FIG9}, except that the
FM(S)-PMS and PMS-AM(S) phase transitions appear to change from first
to second order at a smaller value of $y$. For $\kp_A=0$ it appears to
be second order for all $y$.
This indicates that the point
where the FM(S)-PMS and PMS-AM(S) phase transitions change their order
has a similar $\kp_A$-dependence as the phase transition line in the
ISF model shown in fig.~\ref{FIG3}.

\begin{figure}%[t]
 \centerline {
 \fpsysize=14.0cm
 \fpsbox{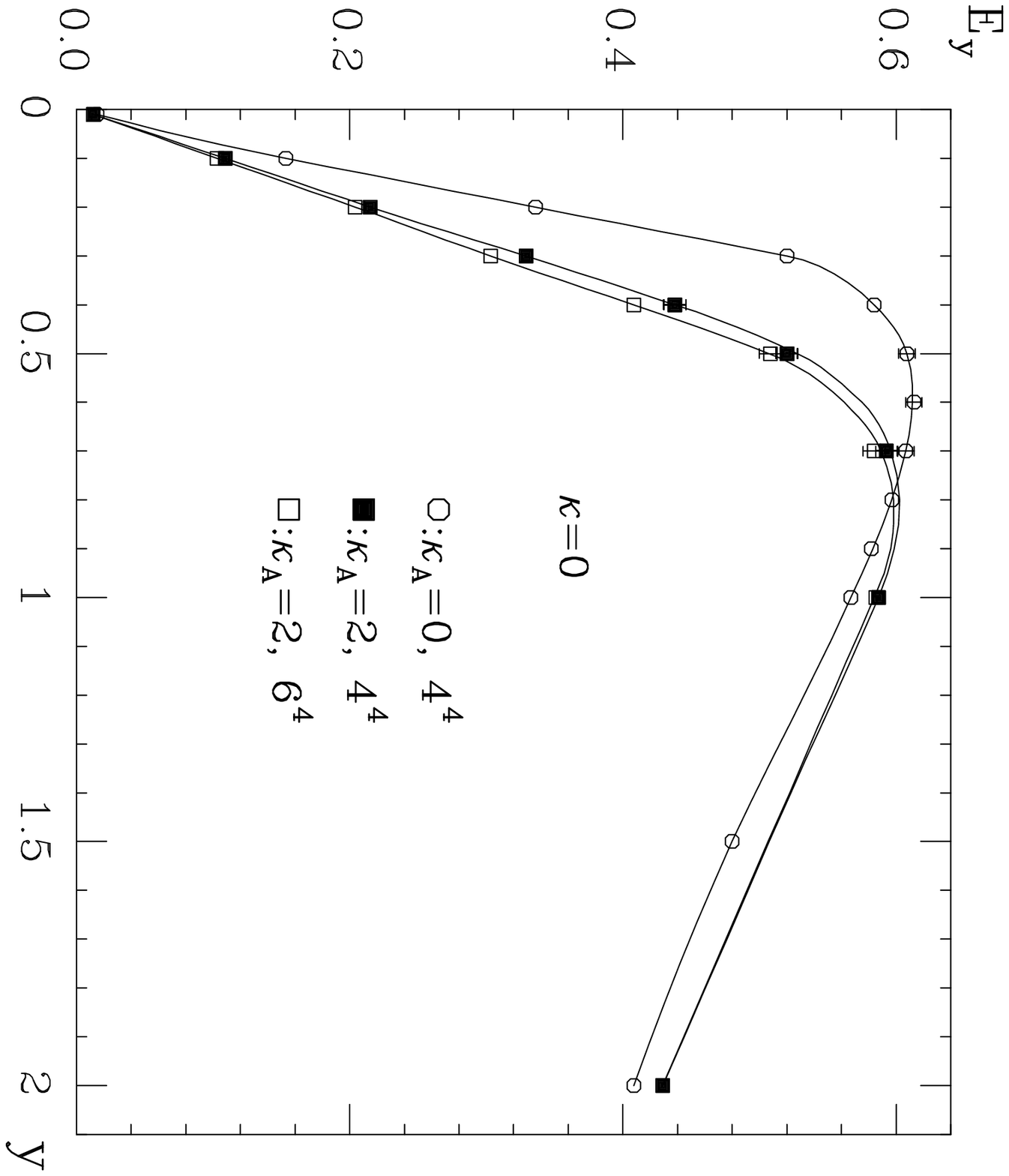}
 }
%\vspace*{12.0cm}
\vspace*{-1.0cm}
\caption{ \noindent {\em The energy $E_{y}$ as a function of $y$ in
the modified NISF  model for $\kp=0$ and $\kp_A=0,2$.
}}
\label{FIG8}
\end{figure}

The
apparent
absence of a phase transition to a PMW phase at small $y$ can be
illustrated from the $y$-dependence of $E_y$.
In fig.~\ref{FIG8} we have displayed the energy $E_y$
as a function of $y$ for two values of $\kp_A$ and $\kp=0$,
on a $4^4$ lattice.
The curve at $\kp_A=0$ (open squares) is identical with the one
shown already in fig.~\ref{FIG2}, for which the PMS phase extends
down to $y=0$.
The maximum of the curve at $\kp_A=2$ (filled squares)
is only shifted a little bit to larger $y$, which indicates that the
$\gm_\mu\gm_5$ term needs a large prefactor $\kp_A$ to become important.
If we try to fix this $\kp_A$ by matching the peaks in fig.~\ref{FIG8}
to those for the ISF model shown in fig.~\ref{FIG2}, we see that
the curve at $\kp_A=2$ of fig.~\ref{FIG8} may be compared with the
curve at $\kp_A=0.3$ in fig.~\ref{FIG2}. This implies that
a relative factor
$\kappa_A \approx 7$ is needed to scale the $\gm_\mu\gm_5$ term in the
modified NISF model to the same strength as in the ISF model.
There is, however, no indication that the curvature
$\partial^2 E_y/\partial y^2$
changes
sign, as for
the curve at $\kp_A=0.3$ in fig.~\ref{FIG2}. Instead the energy appears to
vanish linearly when the value of $y$ is reduced.
This indicates that a phase transition does presumably
also not emerge  when the volume is increased. Also in the other
observables we did not find a sign for a phase transition at non-zero $y$.

%
% FIGURE 8
%
These results show that by the replacement (\eq{SUB}) we have managed to
remove the MS region, but unfortunately we are now left with
a phase diagram with only strong coupling phases.

According to the discussion at the end of sect.~3, the symmetry
restoration and the appearance of a PMW phase is expected to
depend crucially on the dynamics of the low momentum modes.
The small $4^4$ lattice, however, can only accommodate a few low
momentum modes. To improve on this, we
we have included at $\kp_A=2$ also some results on a $6^4$
lattice. All data points lie systematically slightly below the results on
the $4^4$ lattice, but also in this case there is no
indication for a sign change in the curvature at small $y$ which would
reveil the emergence of the
PMW phase. A simulation at larger values of
$\kp_A$ was not possible with our resources of computer time
since the number of CG iterations increases when $\kp_A$
is raised and at the same time the statistics has to be increased
to account for larger autocorrelation times.

%
%
%
% FIGURE 10
%
\begin{figure}%[t]
 \centerline {
 \fpsysize=14.0cm
 \fpsbox{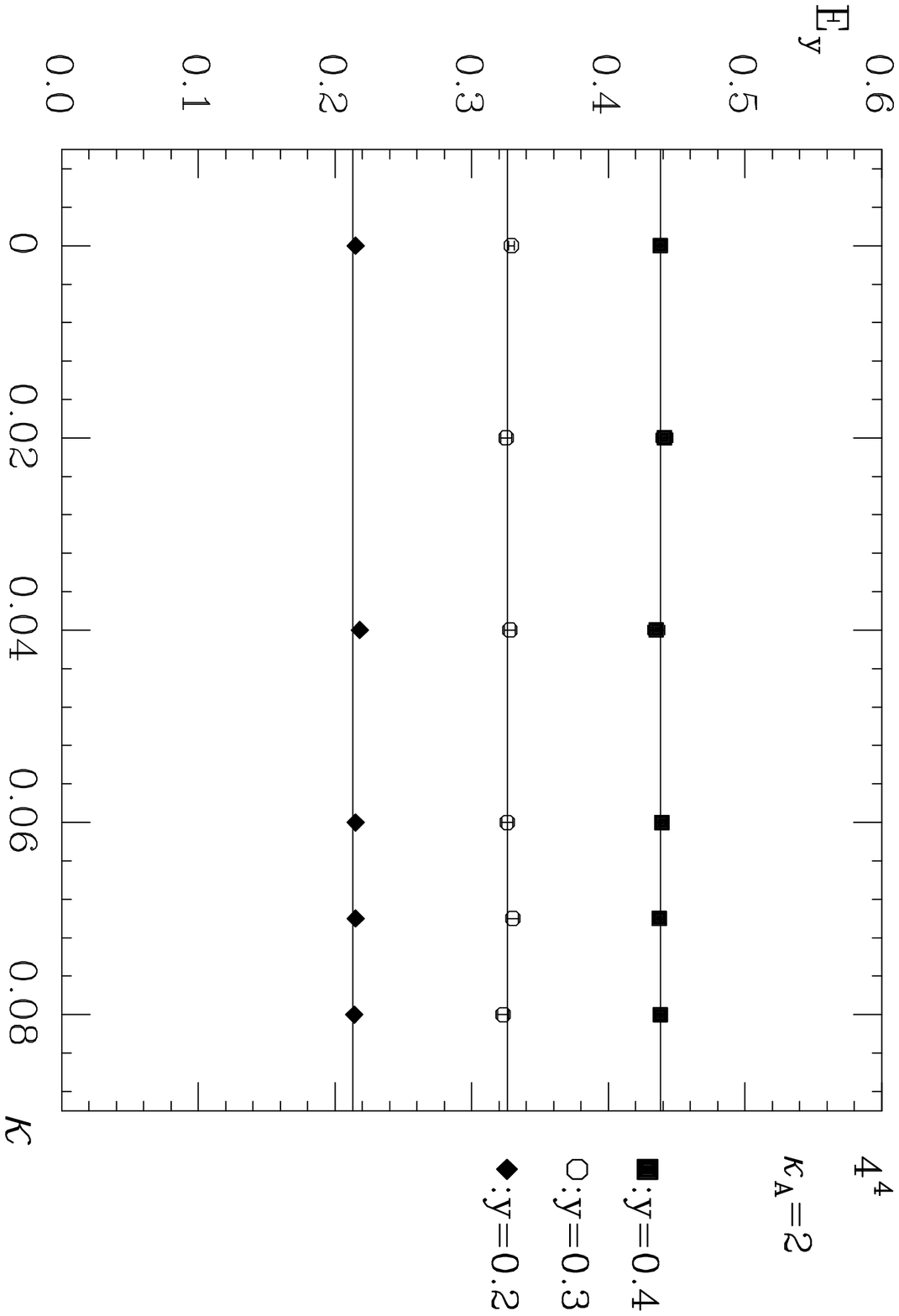}
 }
%\vspace*{12.0cm}
\vspace*{-1.0cm}
\caption{ \noindent {\em The energy $E_{y}$ as a function of $\kp$
for three values of $y$ and $\kp_A=2$ in the modified NISF model.
}}
\label{FIG10}
\end{figure}
Another possibility to
enhance the low
momentum modes of the scalar field,
is to increase the value of $\kp$ towards
a second order
phase transition.
So far we have
considered $\kp=0$ where the scalar
field correlation length is presumably of order one. Therefore the high
momentum components in $V_x$ are very abundant and might prevent
restoration of gauge invariance. It is therefore important to study
the phase structure in a region where the scalar field correlation length
is larger.
In the pure scalar U(1) model the correlation length of the scalar particles
in the PMW diverges when $\kp \nearrow \kp_c^{FM-PM}$.
Also in our model we could hope that by choosing $\kp$ sufficiently
close to $\kp_c$, we can increase the scalar field correlation length
enough that a
PMW phase opens up. As delineated above
the FM(S)-PMS phase transition at small $y$ is however of first order.
This implies
that the correlation length of the scalar particles stays bounded
and our runs at larger $\kp$ show indeed no qualitative
differences from the results for $\kp=0$.
This is illustrated by the $\kp$-dependence of $E_y$.
Interestingly we find that $E_y$ does not depend on $\kp$ for a given
value of $y$ and $\kp < \kp_c^{FM-PM}$.
In fig.~\ref{FIG10} we have monitored the
$\kp$-dependence of $E_y$ for three representative
values of $y$. This shows that a $y$-dependence of $E_y$
as shown in fig.~\ref{FIG8} for $\kp=0$ will also hold for
other $\kp$-values in the PMW phase. A simulation very close
to the FM-PM phase at small $y$ and a precise determination
of the position of the first order transition is hindered
by large hysteresis effects.
Therefore we cannot exclude the possibility that on a much larger
lattice the correlation length very close to $\kp_c$ increases enough
to allow for the appearance of the desired PMW phase.

% FIGURE 11
%
\begin{figure}%[t]
 \centerline {
 \fpsysize=14.0cm
 \fpsbox{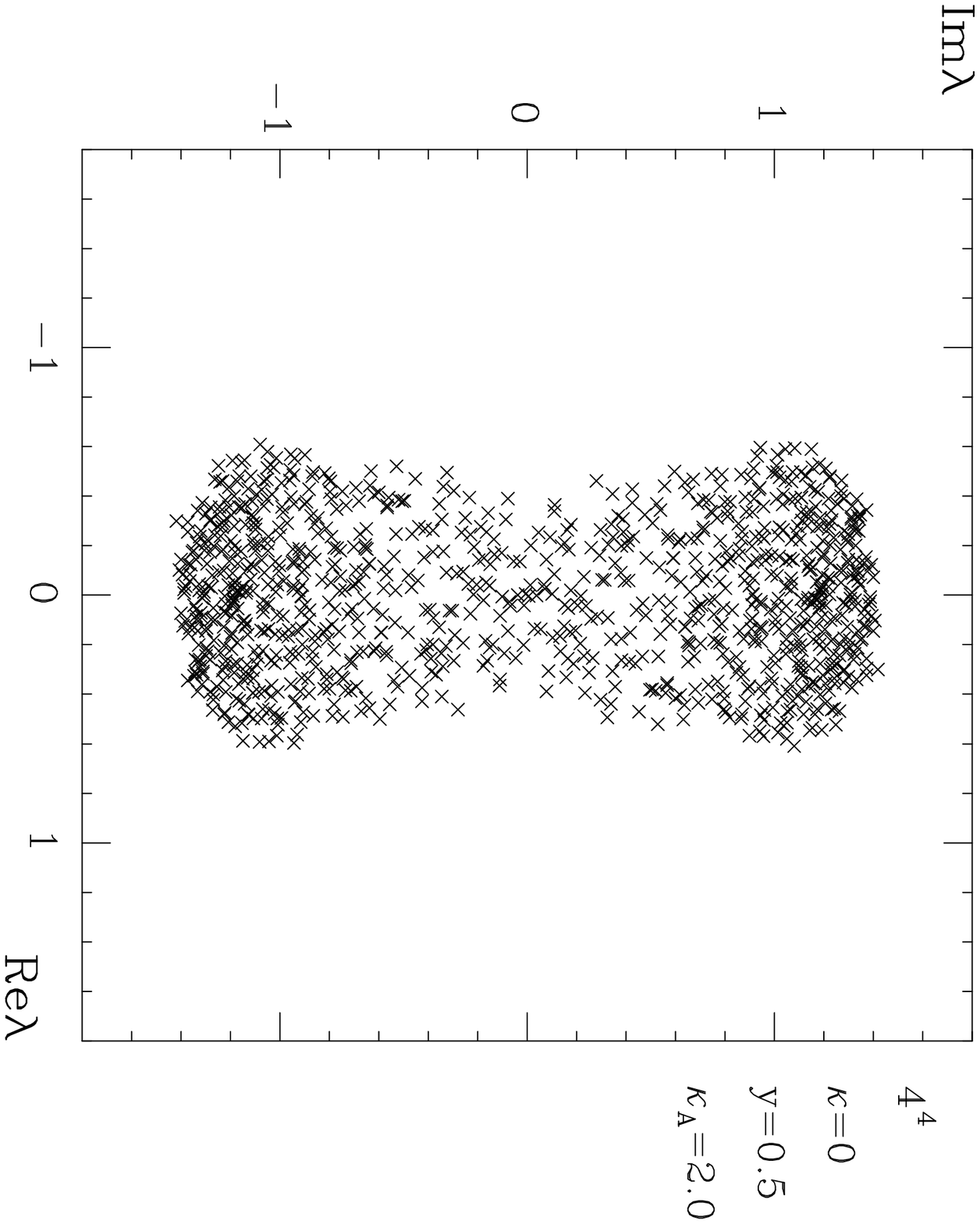}
 }
%\vspace*{12.0cm}
\vspace*{-1.0cm}
\caption{ \noindent {\em
The eigenvalues  of the matrix $M_{NISF}^{off}=M_{NISF}-y \un$ plotted
into the complex plane. The plot contains the eigenvalue spectra for four
independent scalar field configurations which were generated with
the HMCA at $(\kp,y,\kp_A)=(0,0.5,2)$.
The lattice size is $4^4$.
}}
\label{FIG11}
\end{figure}
Additional evidence for
the scenario that there is no
PMW phase emerging for $\kp\nearrow\kp_c$
is obtained from the eigenvalue distributions of the fermion matrix
$M_{NISF}$, modified here by the substitution (\eq{SUB}).
Fig.~\ref{FIG11} contains
the spectra of $M_{NISF}^{off}=M_{NISF}-y \un$
for four independent scalar field configurations, generated
with the HMCA at the point $(\kp,y,\kp_A)=(0,0.5,2)$.
The $M_{NISF}^{off}$ spectra
obtained at other values of $y$ and $\kp$, $\kp_c^{PM-AM}<\kp<\kp_c^{FM-PM}$
with $\kp_A=2$ kept fixed, look similar,
except that for $y \lsim 0.5$ the width  of the distribution on
the real axis vanishes approximately proportionally to $y$.
Fig.~\ref{FIG11} suggests
no formation of a hole inside the distribution which,
as we explained in sect.~6, would be an indication for the
emergence of a PMW phase with restoration of the gauge symmetry  at small $y$.
\section{Final remarks}
We have started our discussion with
examples where restoration of gauge invariance takes place,
the massive Yang-Mills theory and gauge-fermion models,
%where a restoration of gauge symmetry has previously been observed
provided that the symmetry breaking due to the
bare mass term is not too
large. If the mass parameters are raised beyond a critical value
the system is driven to another phase with
scaling
physics differing substantially from that of the original gauge invariant
target model without mass terms. We aimed for a similar dynamical
gauge symmetry restoration in our chiral staggered fermion models, in
which the gauge symmetry is broken by the
high momentum modes.
% of the gauge field.
A strong motivation for this approach is that
it should work without recourse to gauge fixing.

In our approach gauge invariance is restored in the bare action by
integrating over all gauge degrees of freedom
(the `longitudinal components'),
which emerge as dynamical scalar fields in our lattice action,
and which we want to decouple.
We have applied this approach to a U(1) model with axial-vector
couplings,
and tested it in two staggered fermion realizations, denoted as
the NISF models.

The aim was to find in the reduction to external gauge fields, but
keeping the scalars dynamical, a PMW phase (weak
coupling symmetric phase), with massless charged fermions coupled
gauge invariantly to the gauge field, and scalars with masses of
the order of the cut-off.
Our numerical results
indicate, however, no PMW phase in either NISF model.
In the
first
NISF model we
could only find a PMS phase (strong coupling symmetric phase)
connected to large values of the bare coupling
parameter $y$, and a multi-state
region at small $y$ where the
desired gauge symmetry restoration presumably does not take
place.
In the second (`modified') NISF model
our results suggest that
the PMS phase extends down to $y=0$.
One reason for this negative result in the modified NISF model
appears to be the unfortunate
fact that the symmetric phase at small $y$ is separated from
the broken phase by a first order phase transition. Therefore the scalar
field correlation length remains small in the entire symmetric phase and
the symmetry breaking due to the high momentum scalar modes remains
too large.

Nevertheless, we cannot completely
exclude from our numerical results the possibility that at
large values of $\kp_A$,
the renormalization factor of $\gm_5$,
a PMW phase with the desired symmetry restoration could still emerge
in the near vicinity of the FM-PM phase transition.
This may require much larger lattices in order to allow for
sufficient dominance of the low momentum modes in the scaling
region.
It is also possible that after adding
different
counterterms to the action
the first order phase transition between the FM and PM phase at small $y$ may
be changed into a second order one. One could think here of higher
derivative couplings for the scalar field, which suppress the high
momentum modes.
A frustrating fact is, however, that in practice our simple
models fail because of the rapid increase of computational
requirements with volume and with increasing $\kp_A$.

A way out could be the following. To make gauge symmetry restoration
work we need good control over the strength of the symmetry
breaking, which was lacking in our simple models. A natural way
to gain control would be to reduce the scalar field fluctuations
on the scale of the fermionic lattice spacing. We could for
instance use two lattices with  spacings, $a^{\prime}$
for the fermions and $a$
for the bosons, with $a^{\prime} \ll a$.
The bosonic field values on the $a^{\prime}$ lattice are
obtained from the $a$ lattice by interpolation, such that the minimum
wavelength bosonic mode is relatively smooth on the scale of
$a^{\prime}$. The ratio
$a^{\prime}/a$ would then be an important parameter controlling
the strength of gauge symmetry breaking. In the limit $a'/a\ra 0$
this would mean treating the fermions in the continuum (called
`the desperated method' in ref.~\cc{Sm88}). The detailed
implementation, however, of such a program is not so easy (cf.
ref.~\cc{Goe92} for work on interpolating lattice gauge fields).

As an alternative one could start from a
gauge fixed continuum model and regularize it using the lattice and
staggered fermions. The model then becomes very
similar to the Rome proposal \cc{ROME}, which uses Wilson fermions.
A non-perturbative
test of this gauge fixing approach is difficult because
of technical obstacles.
One can still carry
out simpler tests, like quenched simulations
in gauge-fixed background configurations.
Recently we have presented results in this direction
for the case of a U(1) model with axial-vector couplings in two
dimensions \cc{SMOOTH}. \\

\noindent {\bf Acknowledgements}\nopagebreak \\ \nopagebreak
The numerical calculations were performed on the CRAY Y-MP4/464
at SARA, Amsterdam. This research was supported by the ``Stichting voor
Fun\-da\-men\-teel On\-der\-zoek der Materie (FOM)'',
by the ``Stichting Nationale Computer Faciliteiten (NCF)'' and by
the DOE under contract DE-FG03-91ER40546.
\append
In this appendix we
give the derivation of the momentum space relation
(\eq{MSPACE}). The formula is obtained by inserting
\be
\chi_x=\frac{1}{V}\sum_q \sum_b
\exp \left[i(q+b\pi)x \right] \chi_{q+b\pi}
\lb{AP1}
\ee
in the Frourier transform of eq.~(\eq{SDT}),
\be
\Psi(p)=\frac{1}{8} \sum_c \sum_x \exp(-ipx) \gm^{x+c} \chi_{x+c} \;.
\ee
The $\chi_p$ in (\eq{AP1}) is the Fourier transform of $\chi_x$ with
momentum $p=q +b\pi$.  The $\sum_q$ in eq.~(\eq{AP1}) is the
sum over the restricted Brillouin zone,
$-\pi/2 < q_{\mu} \leq +\pi/2$.
After writing $x=y+d$, with $y$ on a lattice with double lattice
distance, and $d$ running over a hyper cube, we obtain,
\be
\Psi(p)=\frac{1}{8V} \sum_q \sum_y\sum_{bcd}
\exp \left[-i(p-q)y \right]
\exp \left[ -i(p-q)d +iqc + ib(c+d)\pi \right] \gm^{c+d} \chi_{q+b\pi} \;,
\ee
where we have
used $\gm^y=1$ and $\exp (i\pi by)=1$.
Next we note that
$\sum_y \exp [i(p-q)y] = \frac{V}{16}\dl_{p,q}$ for $p,q$ in the
restricted Brillouin zone, and use
\bea
\sum_d \exp[i(d+c)b\pi]\gm^{d+c}_{\al\kp} &=& \sum_d
\exp(id b\pi)\gm^d_{\al\kp}\nonumber\\
&\equiv& 8 T_{\al\kp,b}\;,
\eea
to obtain (\eq{MSPACE}) with
\be
Z(p)=\prod_{\mu} \exp( i p_\m/2) \cos (p_{\mu}/2)\;
\ee
(Note that in ref.~\cc{Sm92} a factor $1/2$ is missing
in the expressions for $T$ and $Z(p)$).
By using the formula for irreducible representations of a group
\be
\sum_g {\cal D}^r_{\al\kp}(g){\cal D}^s_{\bt\lm}(g)^* =
\frac{1}{\Tr {\cal D}}\dl_{rs}\dl_{\al\bt}\dl_{\kp\lm}
\ee
with a normalized sum, $\sum_g=1$,
for the case of the fundamental representation of the group of 32 elements
$\pm \gm^d$, and
\be
\sum_b \exp \left[ ib(c-d)\pi\right] = 16\dl_{c,d} \; ,
\ee
it is straightforward to check the unitarity of $T$.
%
%
%
% LITERATURE
%
%

\end{document}